\documentclass[prl,aps, superscriptaddress, twocolumn,showpacs,preprintnumbers,amsmath,amssymb, 10pt]{revtex4}

\usepackage[T1]{fontenc}
\usepackage[utf8]{luainputenc}
\usepackage{amstext}
\usepackage{epsfig}
\usepackage{amssymb,amsmath,wasysym}
\usepackage[normalem]{ulem}  
\usepackage{graphicx}
\usepackage[absolute]{textpos}
\usepackage{subscript}
\usepackage{color}

\newcommand{\beq}{\begin{equation}}
\newcommand{\eeq}{\end{equation}}
\newcommand{\beqarr}{\begin{eqnarray}}
\newcommand{\eeqarr}{\end{eqnarray}}

\newcommand{\footnoteremember}[2]{
\footnote{#2}
\newcounter{#1}
\setcounter{#1}{\value{footnote}}
}
\newcommand{\footnoterecall}[1]{
\footnotemark[\value{#1}]
}

\makeatletter

\begin{document}


\title{Complex Rotating Waves and Long Transients in a Ring-Network of Electrochemical Oscillators with Sparse Random Cross-Connections}
\author{Michael Sebek}
\affiliation{Department of Chemistry, 
Saint Louis University, 
3501 Laclede Ave., St. Louis, Missouri 63103, USA}
\author{Ralf T\"onjes}
\affiliation{Institute of Physics and Astronomy, 
Potsdam University, 
14476 Potsdam-Golm, Germany}
\author{Istv\'{a}n Z. Kiss} 
\affiliation{Department of Chemistry, 
Saint Louis University, 
3501 Laclede Ave., St. Louis, Missouri 63103, USA}
 
\graphicspath{{figures/}} 

\date{\today}


\begin{abstract}
We perform experiments and phase model simulations with a ring network of oscillatory electrochemical reactions to explore the effect of random connections and non-isochronocity of the interactions on the pattern formation. A few additional links facilitate the emergence of the fully synchronized state. 
With larger non-isochronicity, complex rotating waves or persistent irregular phase dynamics can derail the convergence to global synchronization. The observed long transients of irregular phase dynamics exemplify the possibility of a sudden onset of hyper synchronous behavior without any external stimulus or network reorganization.
\end{abstract}

\maketitle

\begin{figure}
\includegraphics[width=0.493\columnwidth]{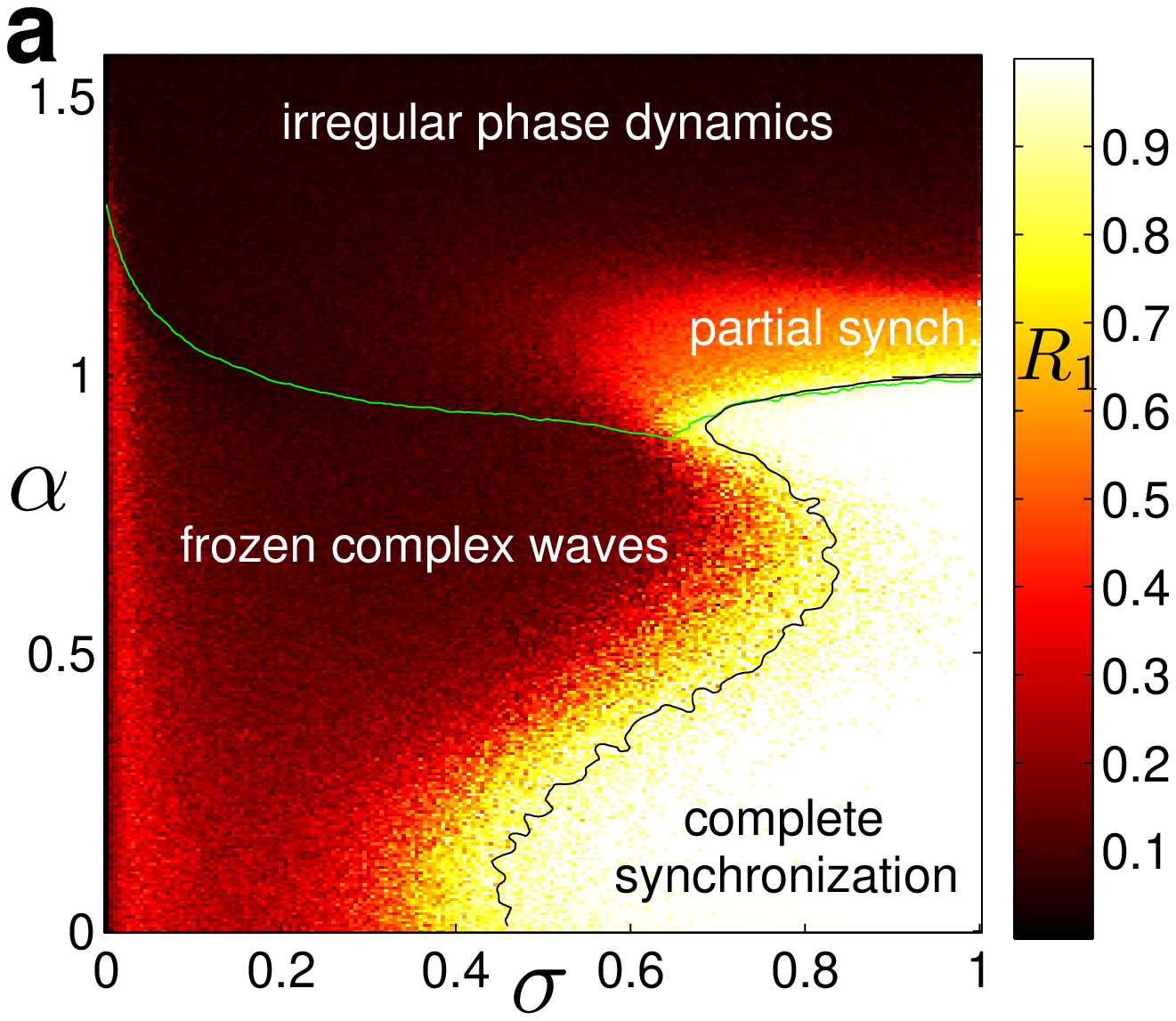}	\label{fig1a}
\includegraphics[width=0.493\columnwidth]{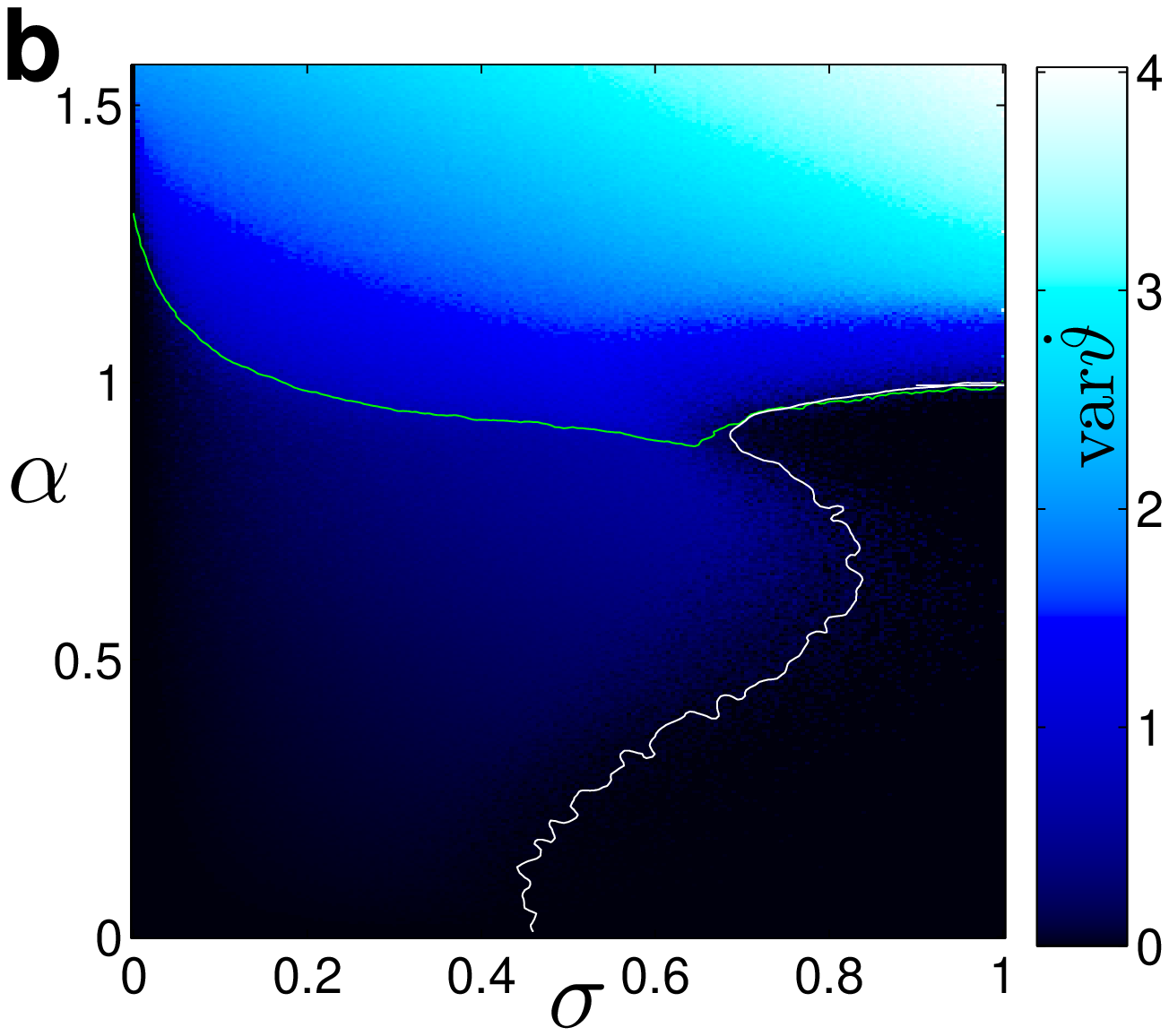}	\label{fig1b}
\includegraphics[width=0.493\columnwidth]{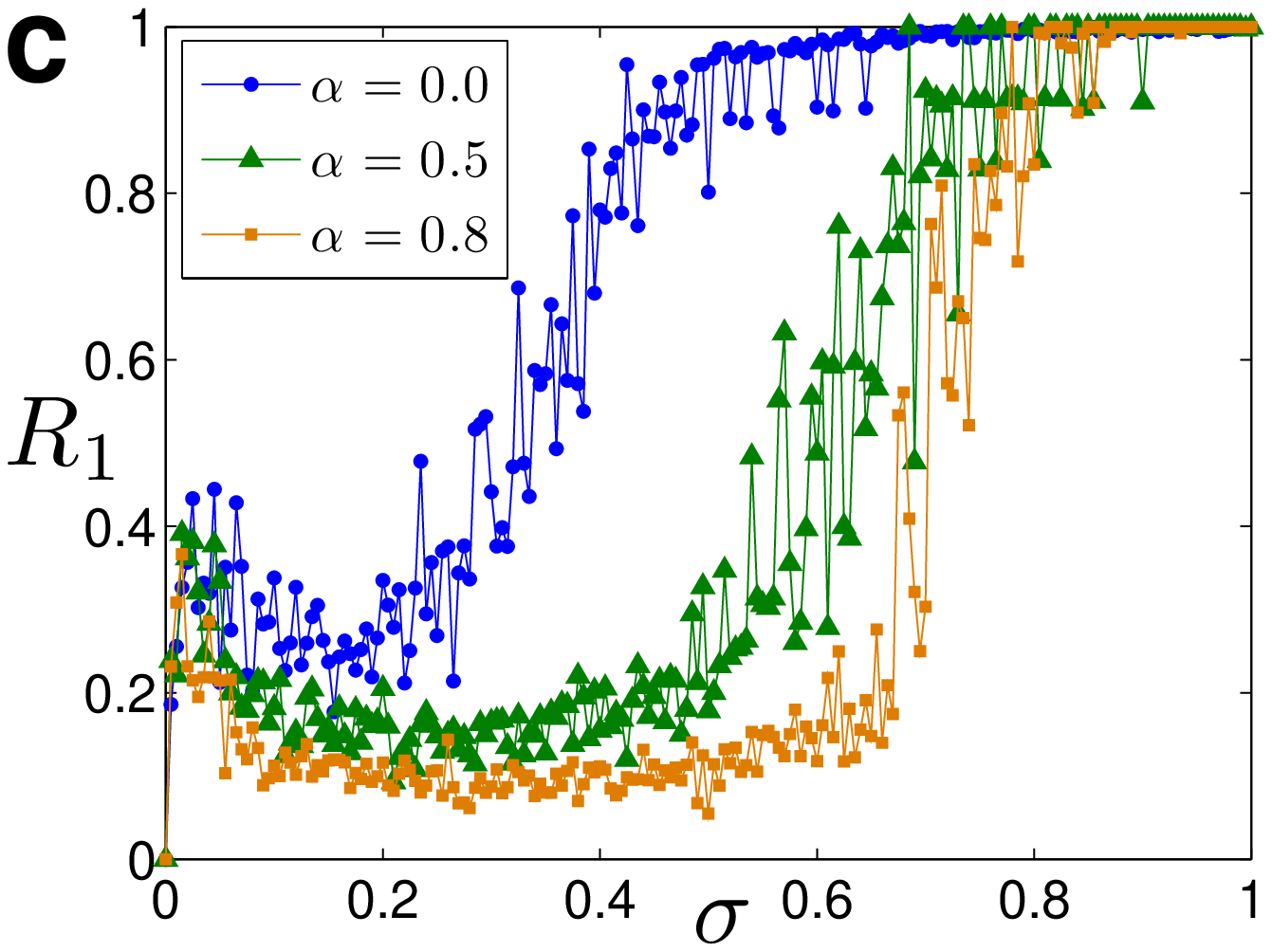}	\label{fig1c}
\includegraphics[width=0.493\columnwidth]{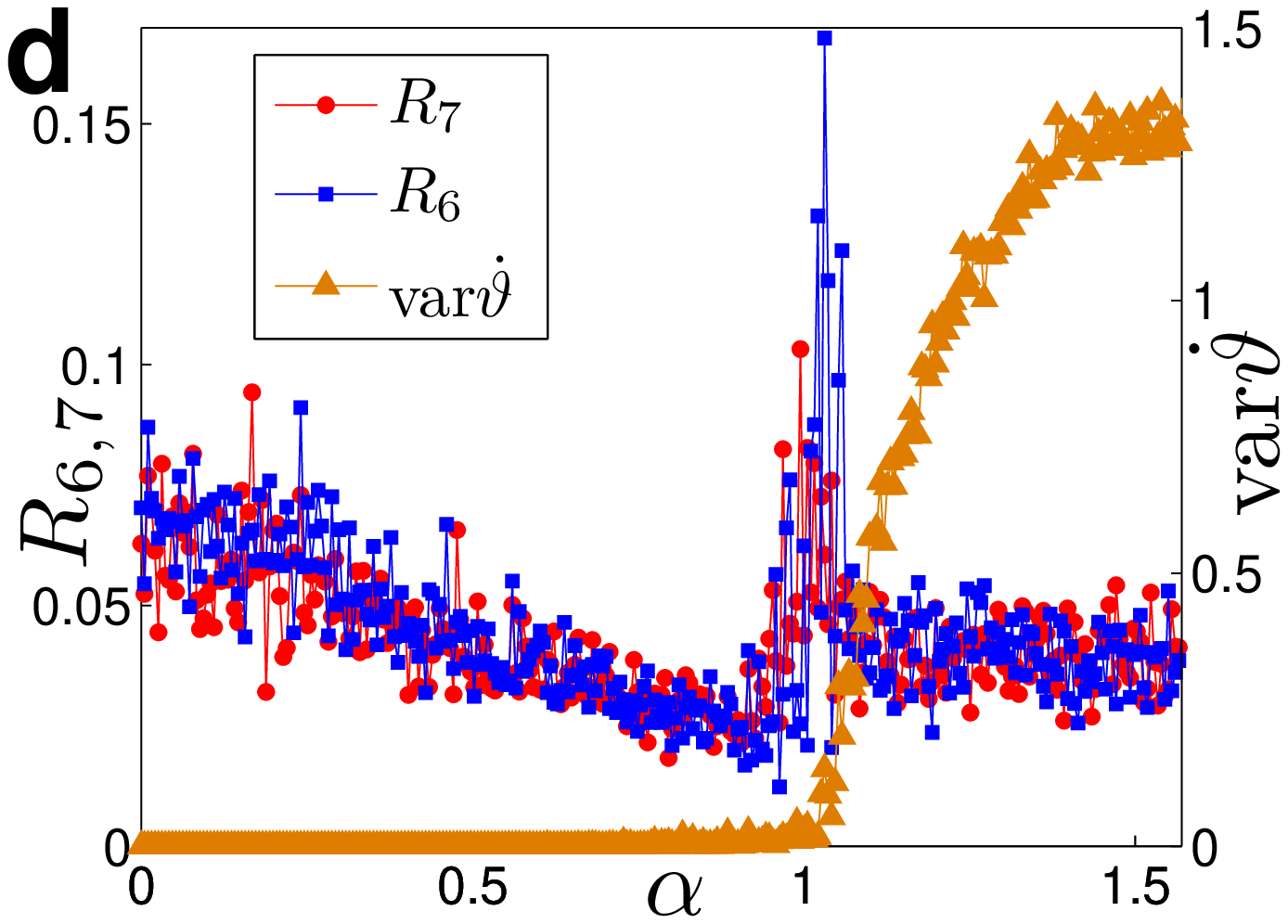}	\label{fig1d}
\caption{(Color online) Order parameters in model (\ref{eq:phasemodel}) at time $t=500$ as functions of shortcut density $\sigma$ and non isochronicity parameter $\alpha$ averaged over ten random network realizations with $N=500$ oscillators and rotating wave initial conditions. {(a)} Color coded Kuramoto order parameter $R_1$ and {(b)} variance $\textrm{var}\dot\vartheta$ of phase velocities. The black and the white line in {(a)} and {(b)}, respectively, mark the contour of $R_1=0.9$. The light (green) line marks the contour line of $\textrm{var}\dot\vartheta=0.2$. {(c)} Kuramoto order parameter as a function of $\sigma$ for three different values of $\alpha$ (cf.{\ref{fig-sync-trans}c}). (d) Mean cluster order parameters $R_6$,$R_7$ and variance of phase velocities as functions of $\alpha$ at $\sigma=0.15$.}
\label{fig-sim-model}
\end{figure}

\begin{figure}
\includegraphics[height=0.3\columnwidth]{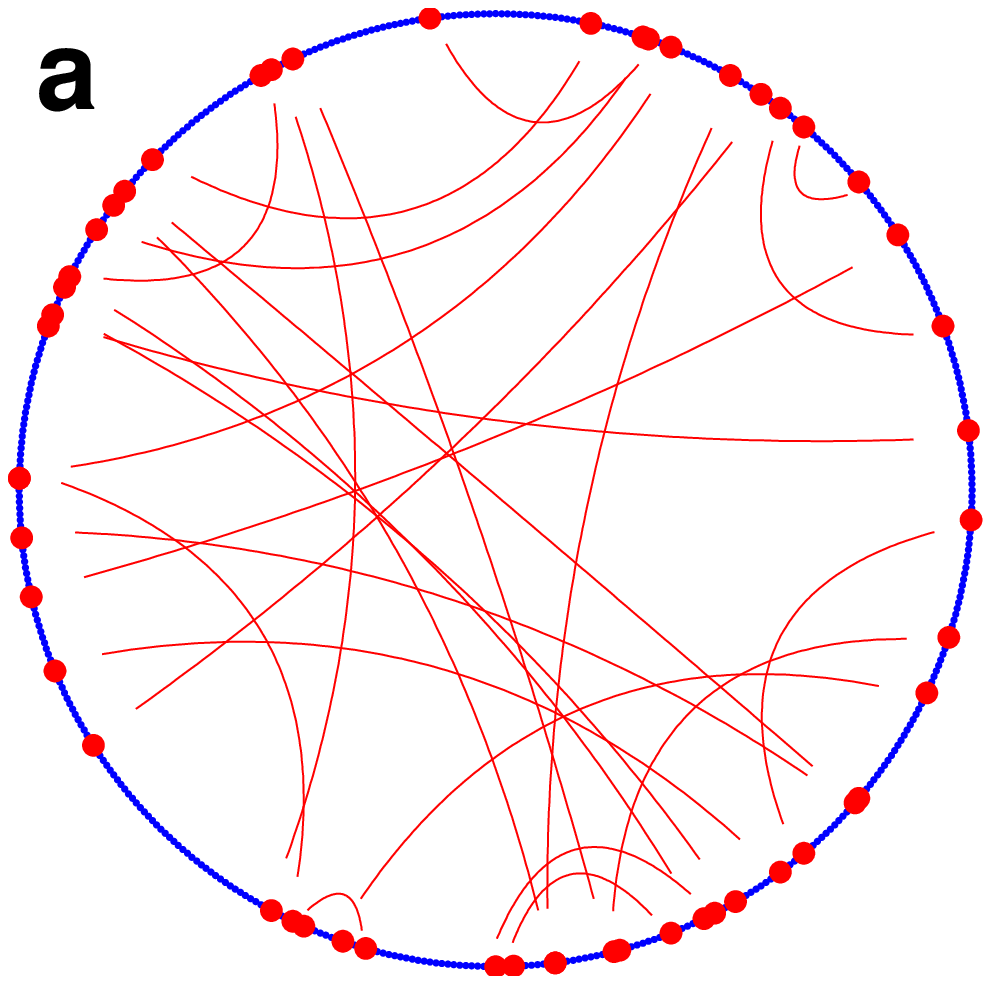}	\label{fig2a}\qquad\quad
\includegraphics[height=0.3\columnwidth]{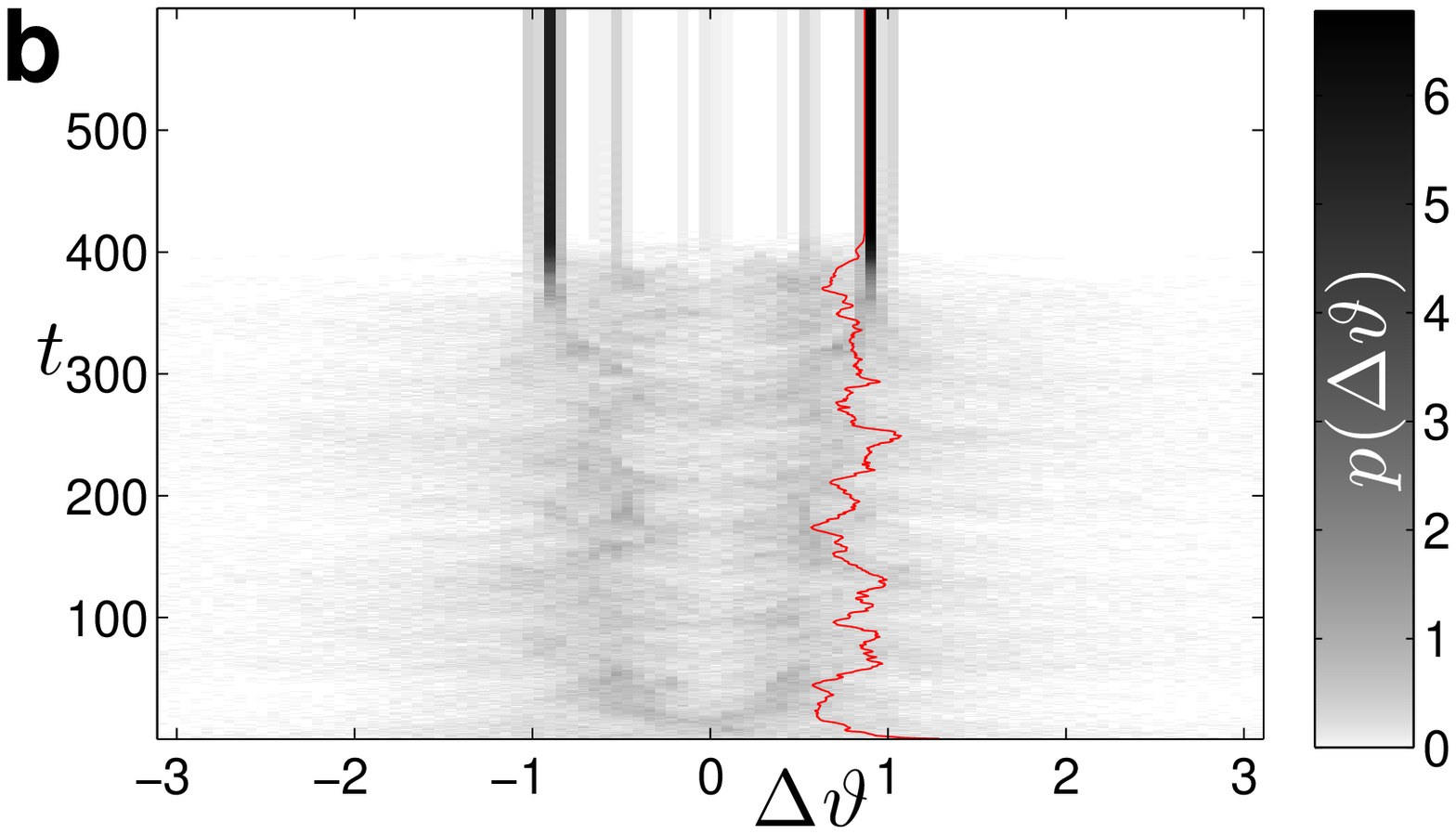}	\label{fig2b}
\includegraphics[width=0.98\columnwidth]{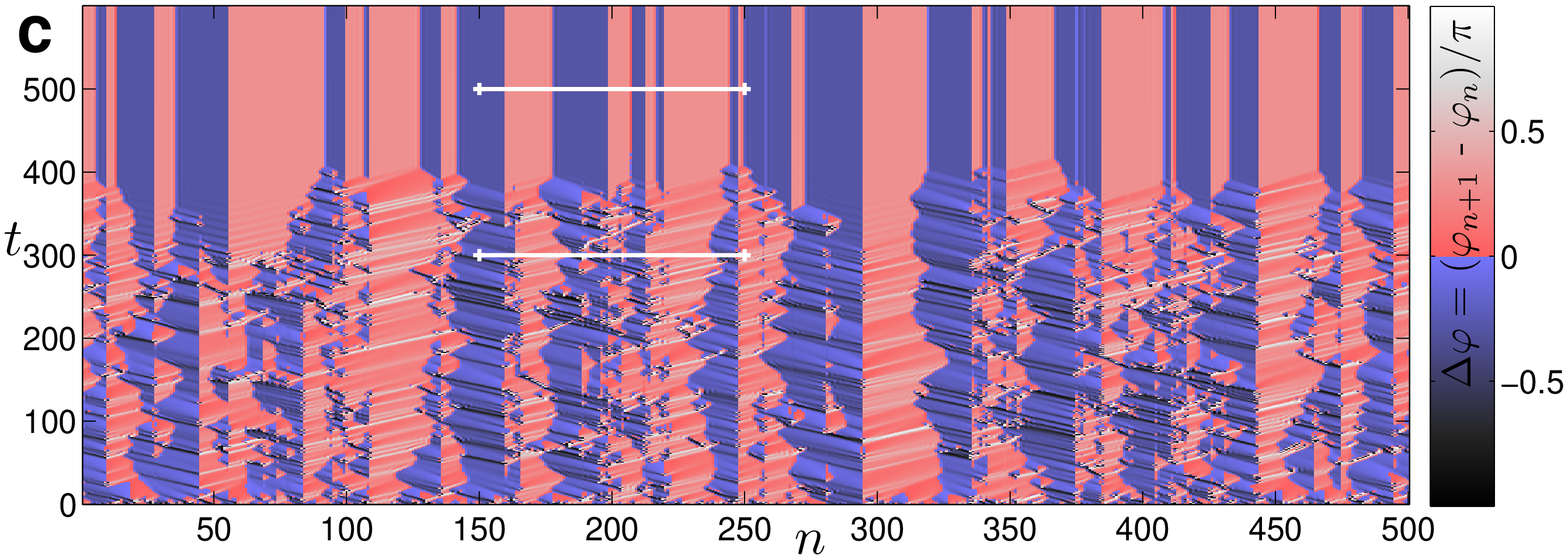}	\label{fig2c}
\includegraphics[width=0.98\columnwidth]{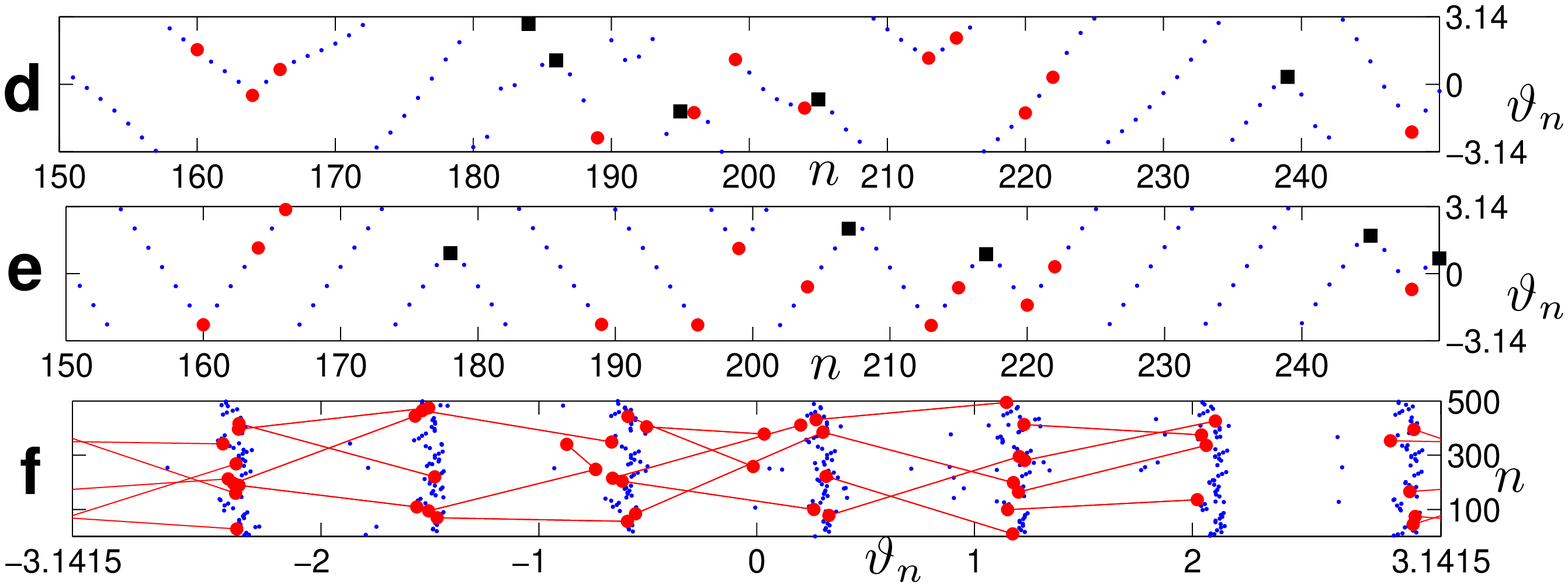}	\label{fig2def}
\caption{(Color online) Long transient to a complex frozen rotating wave pattern from random initial phases in a network of $N$= 500 phase oscillators with $\alpha$= -1.15 and 
shortcut density $\sigma$= 0.05.
{(a)} Ring network 
with 26 additional random shortcuts 
{(b)} Time evolution of the density of next neighbor phase differences. The solid (red) line marks the average of $|\Delta\vartheta|$.
{(c)} Time evolution of next neighbor phase differences modulus $\pi$ (color coded). Dark (blue) colors indicate phase waves to the 
right and light (red) colors indicate phase waves to the left.  The white lines indicate phase profiles $\vartheta_{n}$ between 150 $\leq$ n $\leq$ 250 shown in 
sub figures (d) at $t$= 300 and (e) at $t$= 500. 
Large (red) circles and solid (red) lines in panels (a,d-f) indicate nodes with shortcut connections.
Black squares in panels (d,e) indicate dynamically realized centers of phase waves. Panel (f) shows clustering of the phases at $t$=500 with $R_7\approx 0.9$ and cross-links (solid lines) connecting neighboring clusters.}
\label{fig-sim-turb}
\end{figure}

\begin{figure}
\begin{center}
\includegraphics[width=0.45\columnwidth]{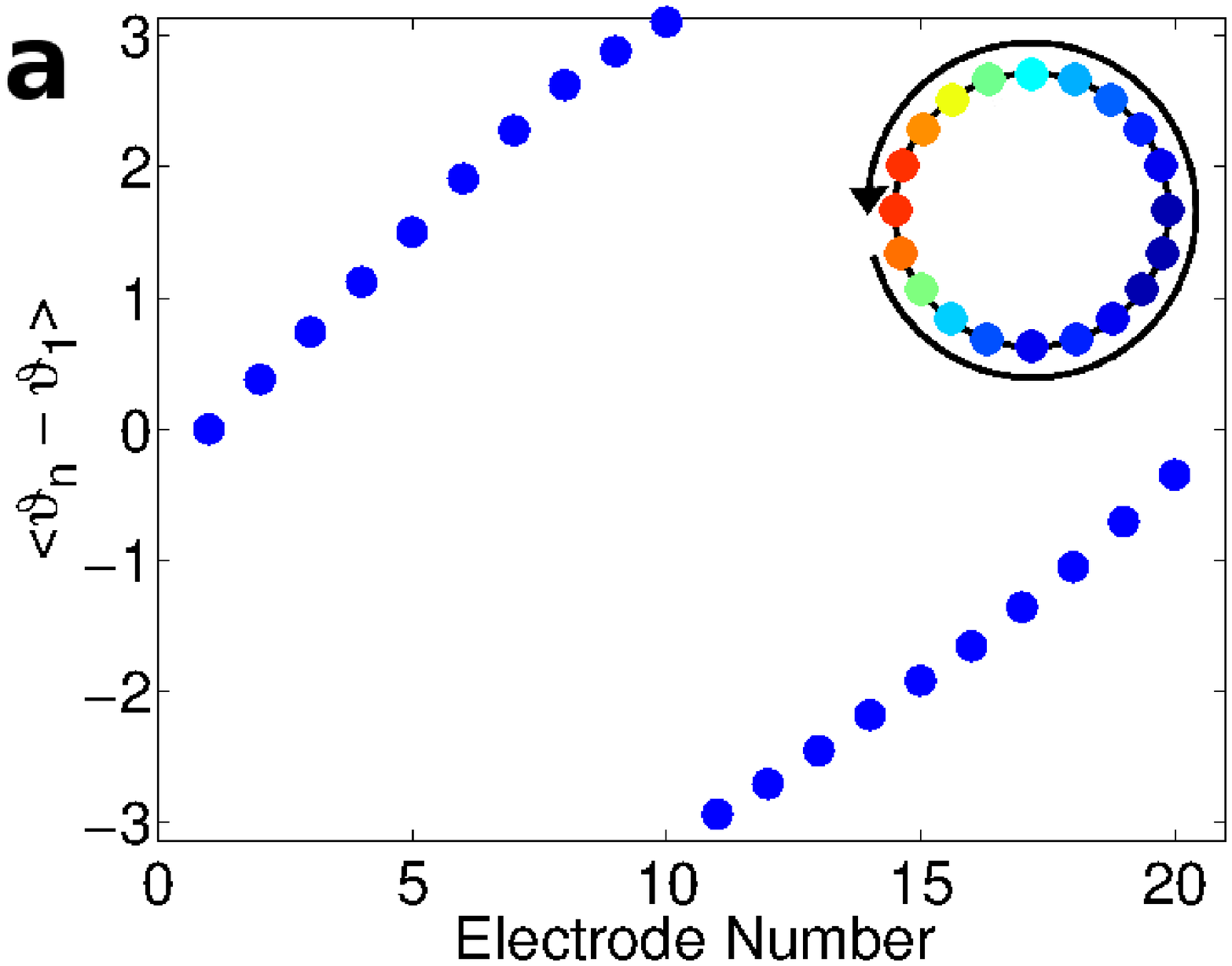}{\qquad}	\label{fig3a}
\includegraphics[width=0.45\columnwidth]{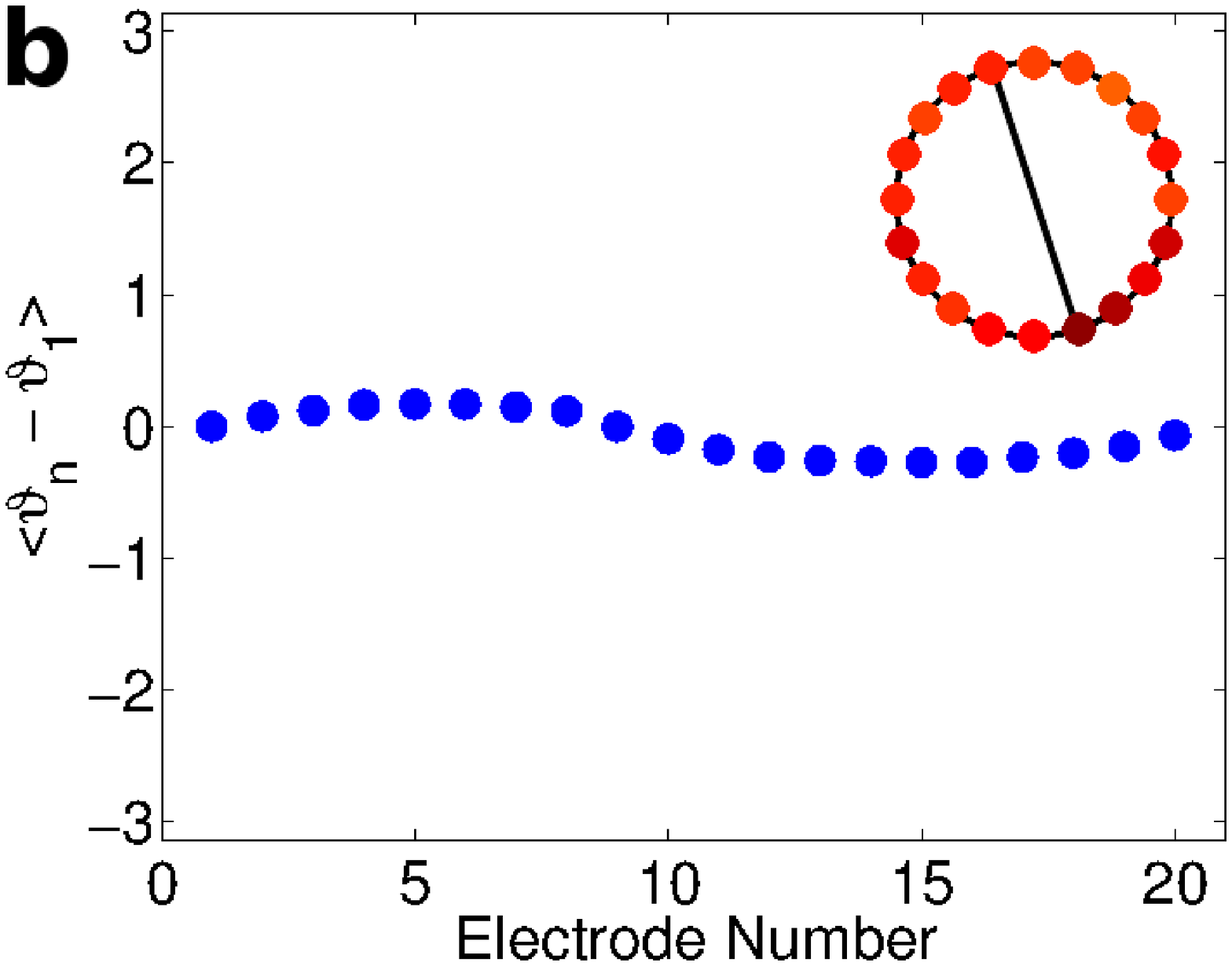}	\label{fig3b}
\includegraphics[width=0.45\columnwidth]{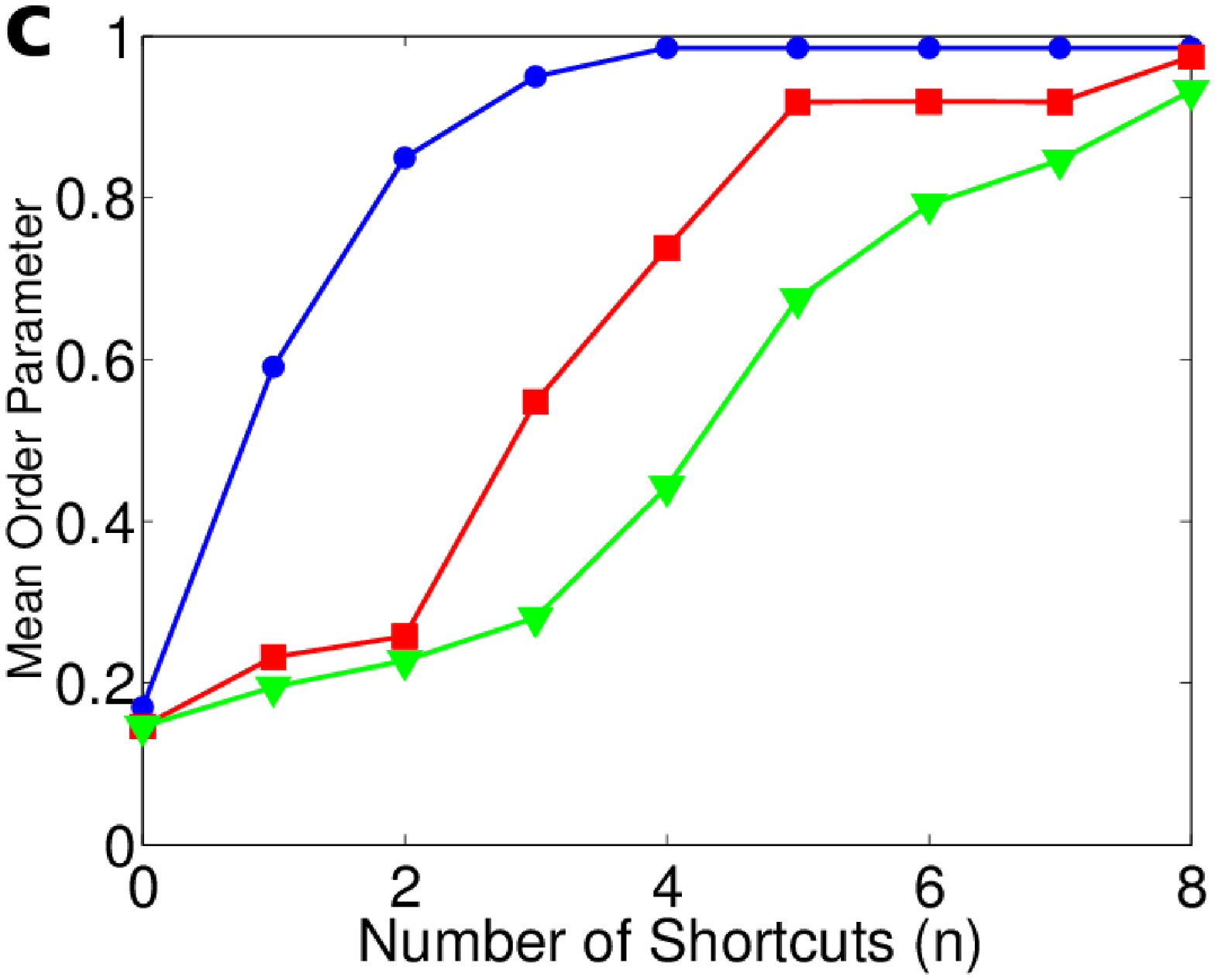}{\qquad}	\label{fig3c}
\includegraphics[width=0.45\columnwidth]{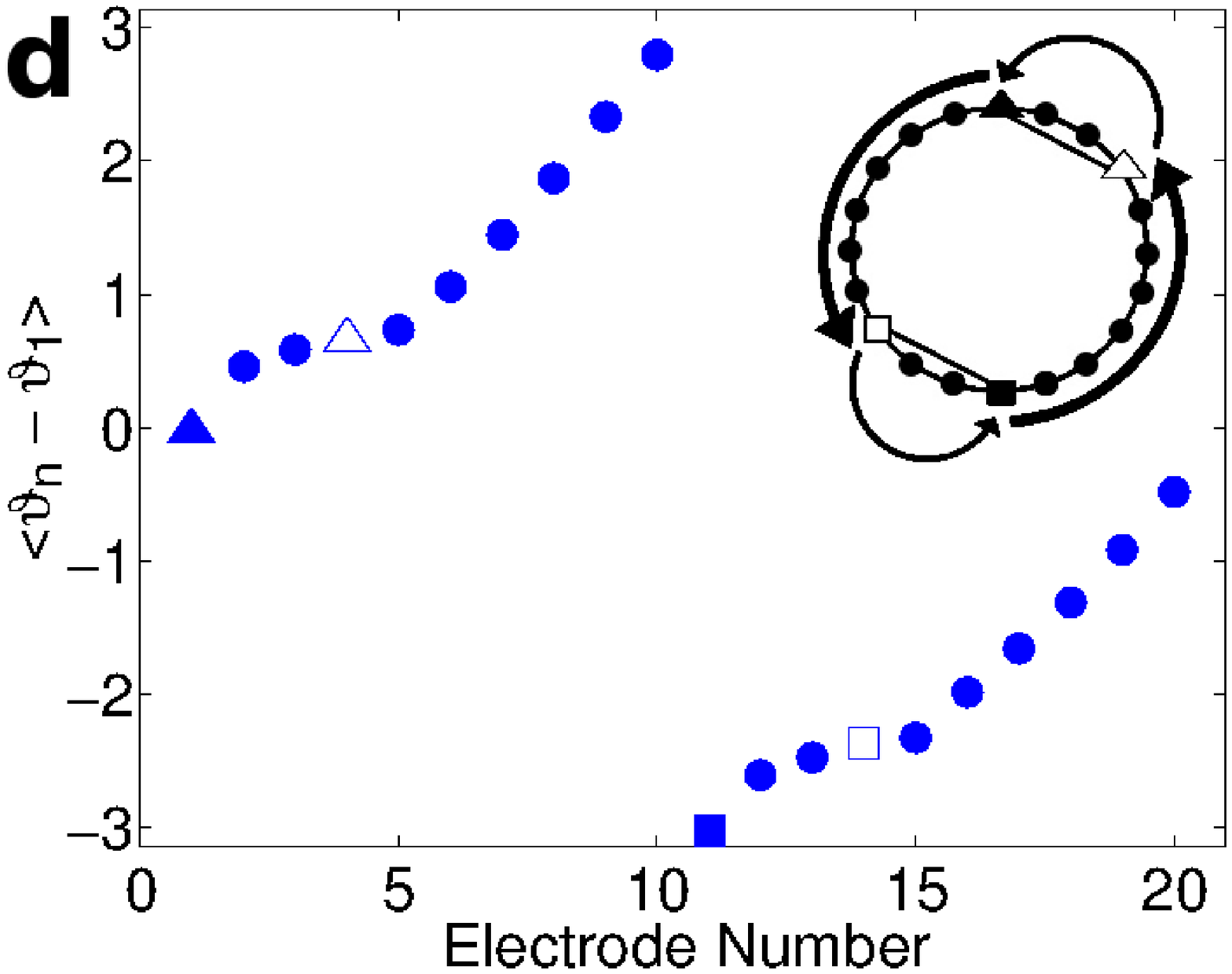}	\label{fig3d}
\includegraphics[width=0.45\columnwidth]{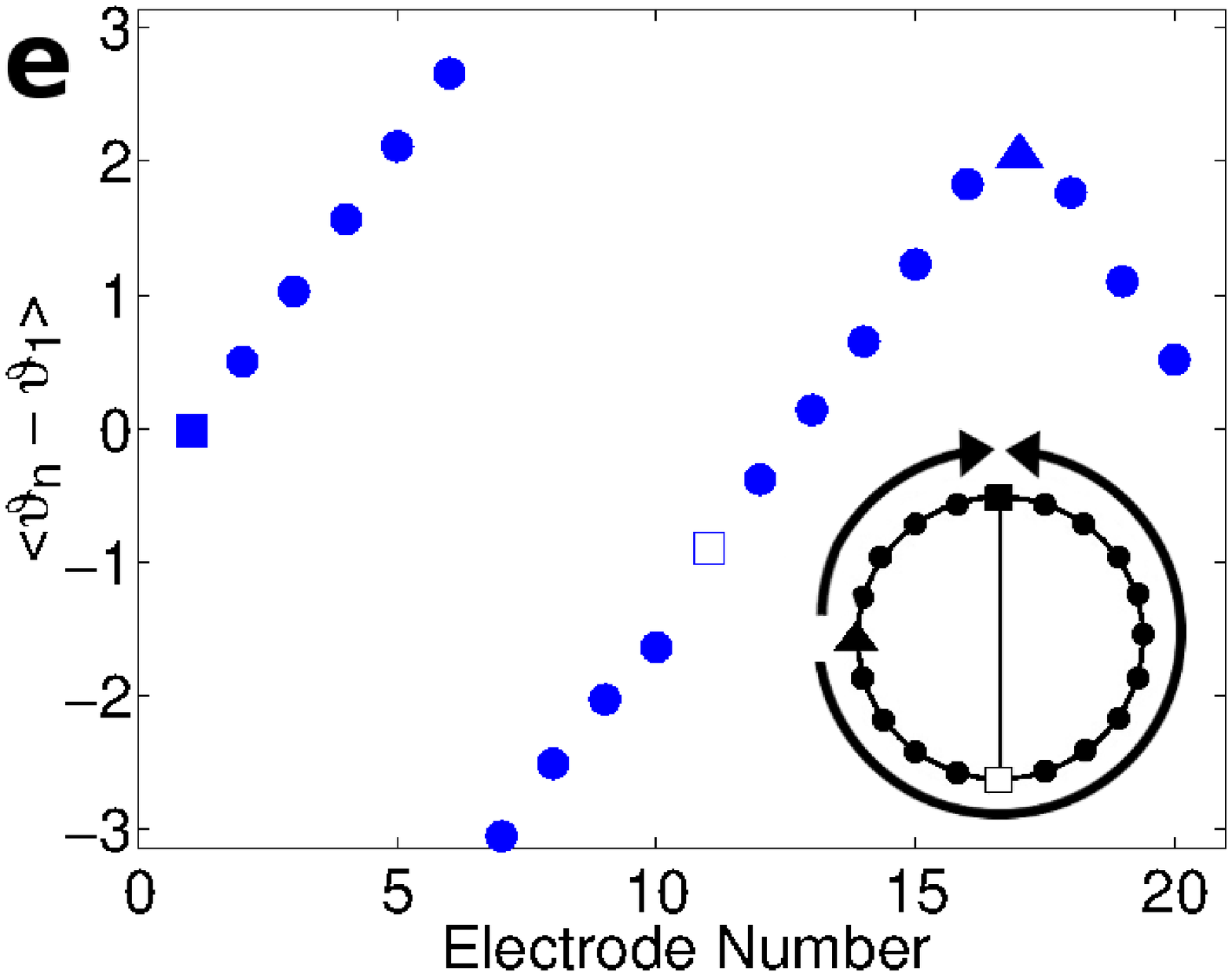}{\qquad}	\label{fig3e}
\includegraphics[width=0.45\columnwidth]{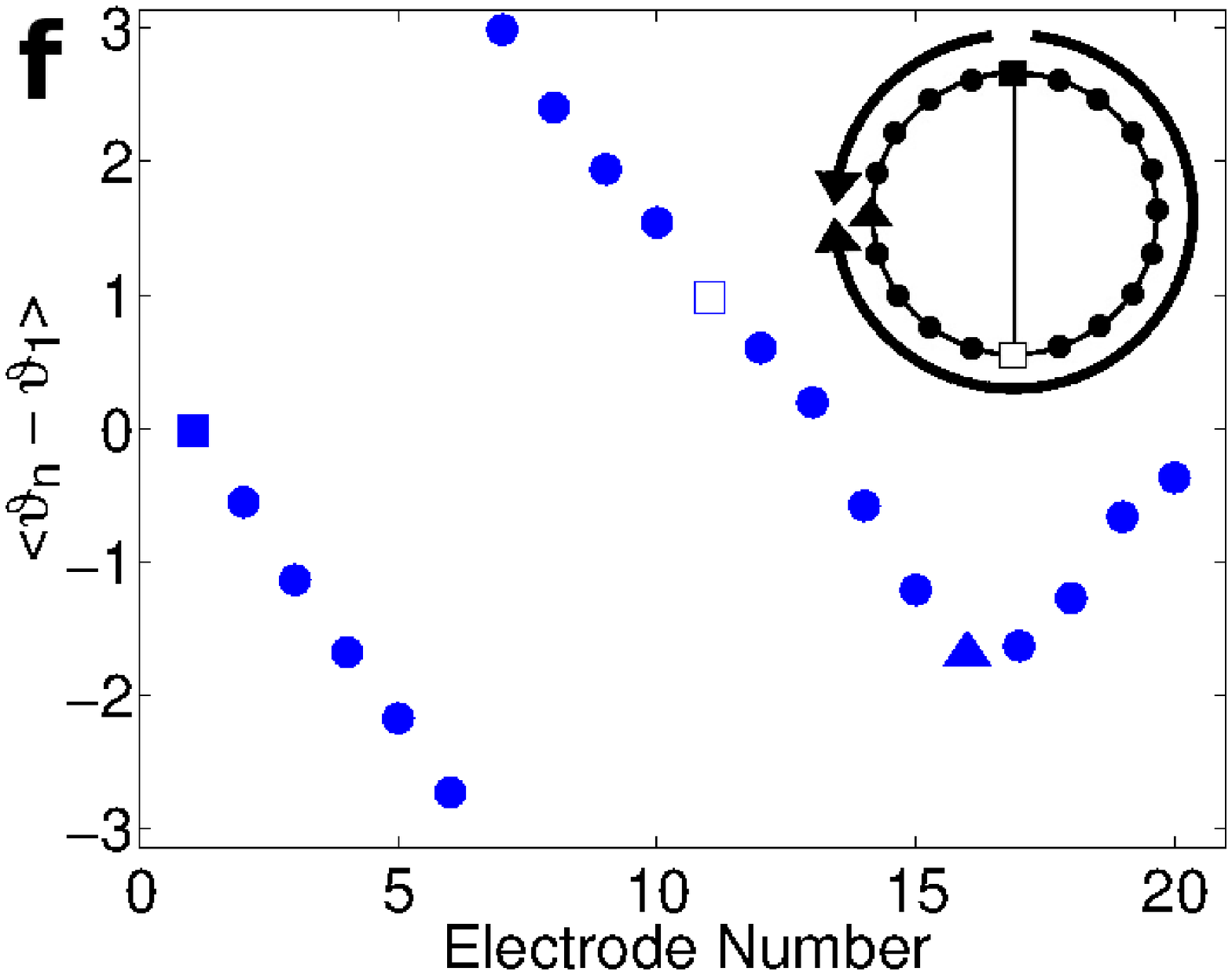}	\label{fig3f}
\end{center}
\caption{The impact of shortcuts and $\alpha$ on the formation of complex rotating waves and on synchronization.  
{(a)} Initial rotating wave ($V$=1105 mV, $K$=0.20 mS, $\alpha$=0). 
{(b)} Synchrony induced by one long distance shortcut. 
{(c)} The mean order parameter with increasing number of shortcuts at $\alpha=0$ (circles), -0.97 (squares) and -1.3 (triangles). 
$V$=1110 mV, $C\textsubscript{c}$=82 $\mu$F, $K$=0.10 mS at $\alpha$=-0.97 and $K$=0.025 mS at $\alpha$=-1.3. 
{(d)} Short distance shortcuts at $\alpha$=0 yield jumping waves ($V$=1110 mV). 
{(e)} Long distance shortcut at $\alpha$=-0.97 yields a frozen complex rotating  pattern with a source (triangle) away from the heterogeneity ($V$=1110 mV). 
{(f)} Long distance shortcut at $\alpha$=0.76 yields a frozen complex rotating  pattern with a source (full square) on the heterogeneity ($V$=1245 mV, 
$C\textsubscript{ind}$=1 mF, $K$=0.40 mS).}
\label{fig-sync-trans}
\end{figure}

%
%
Wave propagation of activity of oscillatory units in rings or linear
chains is a fundamental type of pattern formation, that occurs in many
biological systems, e.g., motion of leach \cite{Iwasaki:2014:978-983},
the segmentation clock \cite{Lauschke:2012:101-105}, or brain
wave activities in the cortex \cite{Ermentrout:2001:33-44}. A rotating
pinwheel was one of the first type of chemical pattern formation identified
in the BZ reaction on a ring \cite{Noszticzius:1987:619-620}. As a ring geometry is often used
in chemistry, rotating phase wave patterns have been observed in
a large number of systems, e.g., in electrochemical reactions \cite{Varela:2005:2429-2439},
heterogeneous catalysis \cite{Luss:2005:254-274}, coupled BZ reactors \cite{Laplante:1992:4931-4934} and
micro droplets \cite{Tompkins:2014:4397-4402}. Mathematical analysis
using phase models interpreted the existence and 
local stability of rotating waves in ring networks \cite{Ermentrout:1985:55-74,Ermentrout:1992:1665-1687,Kopell:1986:623-660}.
It was found that the fully synchronized, zero phase lag, non-rotating state is the most attracting
solution, locally and globally. However, with increasing system size, rotating waves with higher winding number become more
probable in the aggregate \cite{Wiley:2006:015103}.
\\
Complex engineered and biological systems can often be described as networks of discrete, interacting units \cite{Albert:2002:47-97}.  Considering the 
prevalence of phase waves on rings and chains, a fundamental question is how
the rotating waves manifest in networks that are composed of a regular
ring backbone with a few additional random connections. 
Numerical simulations with phase models on 
sparse directed networks with random initial phases have shown that for sufficiently non-isochronous oscillations, while the fully synchronized state is 
locally stable, persistent irregular phase dynamics is the typically observed behavior \cite{Tonjes:2010:033108}. Such prolonged transient behavior can 
severely impact system response when robust
synchronization is required as it was demonstrated with power grid
models \cite{Menck:2014:3969} or when synchronization is undesirable, e.g., in hyper synchronous neuronal discharges during seizures 
\cite{UhlSing2006}. 
\\
In this paper, we explore the type of spatiotemporal patterns that can be
obtained  with oscillatory chemical reactions
on bidirectional ring networks with random long range connections. The experimental
work is motivated by phase model calculations that predict the presence
of complex rotating waves and long transients in small random networks
with sufficiently non-isochronous oscillations.  The experimental conditions 
allow the analysis of the dependence of pattern formation on the randomness of the network topology and the level of
non-isochronicity of the interactions among the units. 
\\
%
%
%
To study the properties of complex rotating waves on networks we consider weakly coupled, identical limit cycle oscillators with a Kuramoto type phase 
model \cite{Kuramoto:1984} for phase differences in a co-rotating frame of reference
\begin{equation}
\dot \vartheta_n = \sum_{m=1}^N A_{nm} g(\vartheta_m-\vartheta_n). 
\label{eq:phasemodel}
\end{equation}
where $A_{nm}$ represents a coupling matrix and $g(\Delta\vartheta)$ is the average effect of the coupling for oscillators with 
phase difference $\Delta\vartheta$. 
A phase attractive coupling is assumed with  interaction function  $g(\Delta\vartheta) = \sin(\Delta\vartheta-\alpha) + \sin(\alpha)$.  The phase shift 
parameter $\alpha$ is an important system property determined by the average shear flow near the limit cycle in the direction of perturbation caused by 
coupling \cite{Kuramoto:1984}, i.e. $\alpha$ quantifies the non-isochronicity of the oscillations induced by interactions. We note that since the dynamics 
of the model equations (Eq.\ref{eq:phasemodel}) is invariant under a 
change of $\alpha\to -\alpha$, $\vartheta\to -\vartheta$ and $t\to -t$, the phase differences and the frequency shift are inverted when $\alpha$ 
changes sign, such that sources become sinks of rotating waves and vice versa.
We assume non-normalized, bidirectional coupling $A_{mn}=A_{nm}\in\{0,1\}$ on ring networks of $N=500$ oscillators with $N_{sc}=\sigma N$ 
additional random bidirectional links. The initial conditions for the simulations and the experiments is a rotating wave on the ring. (Random initial conditions give comparable results.)
\\
Due to phase attractive coupling  $g'(0)=\cos\alpha>0$ and complete connectedness of the network, 
the fully synchronized, one-cluster state is always a linearly stable solution of (\ref{eq:phasemodel}) \cite{Tonjes:2010:033108}. 
However, the typical behavior of the network starting from globally desynchronized initial conditions, is far more complex than the intuitively expected 
relaxation to the one-cluster state.  We characterize the state of the system by different order parameters. The $k$-cluster order parameters $R_k = 
\left\langle\exp(ik\vartheta_n)\right\rangle$ where the average is taken instantaneously over all oscillators, measure the coherence of the distribution of phases into $k$ 
evenly spaced clusters. The variance $\textrm{var}\dot\vartheta$ of the phase velocities is a measure for frequency synchronization. These ensemble 
averaged measures are shown 
in the $\sigma$ vs. $\alpha$ parameter plane in Figs.~\ref{fig-sim-model}a,b. 
\\
With small $\sigma$, the original ring is divided into linear 
segments between the end points of shortcuts, which can support traveling phase waves.  At the interfaces where two such traveling waves meet, 
the phase differences in a state of stable synchronization are restricted. At low non-isochronicity and low shortcut density these interfaces can be frozen when all topological boundary conditions can be met simultaneously. We refer to such a pattern, which does not change in time, as frozen complex  
rotating wave pattern. When the topological boundary conditions are not met (this is likely to occur with a large number 
of oscillators), slowly changing interfaces are obtained, reminiscent of vortex glasses in 2d 
oscillatory media \cite{Brito:2003:068301}. Both the Kuramoto order parameter $R_1$ and the variance of the phase velocities are small in this regime. 
When the shortcut density is increased,  there exists a topological cross-over to a random network without linear chain segments.
Therefore, when $\sigma$ is increased the system cannot maintain rotating waves and the one-cluster state becomes globally attractive
with $R_1 \approx 1$. 
When $\alpha$ is increased from zero, higher shortcut densities are required for complete synchronization (Figs.~\ref{fig-sim-model}a-c).
\\
At large values of non-isochronicity ($\alpha\gtrsim 1$) a qualitatively 
different type of behavior exists. The topological boundary conditions for rotating phase waves along the 
ring segments with stationary phase differences are very difficult to satisfy simultaneously. 
Instead, the dominant behavior is  persistent irregular dynamics with nonzero $\textrm{var}\dot\vartheta$. 
The distribution of phase differences during the transient and in frozen complex rotating  patterns becomes bimodal, suggesting a preferred phase difference that depends on $\alpha$ 
(Fig.~\ref{fig-sim-turb}b). 
The emergence of persistent 
irregular dynamics is demonstrated in Fig.~\ref{fig-sim-model}d by fixing the shortcut density $\sigma$, and increasing the value of 
$\alpha$.  At the transition between frozen and unfrozen complex rotating  patterns global clustering can arise resulting in a sharp increase in the 
order parameters $R_6$ or $R_7$. This global order is mediated by the end points of the shortcuts in the network (Fig.~\ref{fig-sim-turb}f) : Due to the 
narrow 
distribution of phase differences in a phase locked state, 
oscillators at the same distance to a crosslinked node have the same phase. Global clustering is only observable in a narrow parameter region
at criticality and after a long, system size dependent transient. 
\\
In addition, around the transition point, before a complex rotating wave pattern becomes frozen, very long  transient dynamics 
can be observed.  Figure \ref{fig-sim-turb} shows an example of such transient dynamics, from random initial conditions and with negative non-
isochronicity. The time evolution of next-neighbor phase differences demonstrates the competition between different phase patterns, with stationary 
sinks of the phase waves located at the network heterogeneities and dynamically rearranging sources which may form or annihilate upon collision with a 
sink or at phase slip events. Figures \ref{fig-sim-turb}d,e show a transient and a stationary phase profile, respectively and Fig.\ref{fig-sim-turb}b illustrates 
the time evolution of the phase difference distribution. The phase differences in the stationary phase pattern are peaked sharply around $2\pi/7$ 
resulting in a very precise wavelength and the formation of 7 global phase clusters.
%
%
As shown in Fig.~\ref{fig-sim-model}a, there also exists a narrow regime of partial synchronization for larger shortcut densities which is replaced in a 
sharp discontinuous transition by persistent incoherent phase dynamics at values of $\alpha$ approaching $\pi/2$ and which may be analyzed in a mean 
field approach \cite{Ko:2008:016203}.\\
\noindent 
To confirm the modeling results, experiments were performed with an array of $N=20$, 1.00 mm diameter nickel
wires on which an oscillatory metal dissolution reaction takes place measured by currents. Numerical simulations 
indicate \footnoteremember{Note1}{Supplemental Material}  that regions of frozen rotating patterns, complete synchronization, and irregular phase dynamics can be clearly 
distinguished even in such a small setup. The electrodes are coupled into a ring topology with additional random 
cross-connection via resistances and capacitances.
We report the conductance accross the coupling resistance as coupling strength $K$.
Capacitance is used to introduce non-isochronicity through a phase shift in the coupling current \cite{Wickramasinghe:2013:062911}\footnoterecall{Note1}. 
The initial condition of the experiment is a rotating wave as shown 
in Fig.~\ref{fig-sync-trans}a. First we describe the results with $\alpha$=0, i.e., resistive coupling. 
When a random cross connection was added, one of two scenarios occurred. If the random 
shortcut connected two elements at a distance larger than 4 units, the system 
quickly converged to a fully synchronized state similar to that shown in Fig.~\ref{fig-sync-trans}b. 
When the distance between the shortcut elements was smaller, the rotating waves 
jumped across the connection. Figure \ref{fig-sync-trans}d shows such pattern with two cross connections. \\
%
%
\begin{figure}
\begin{center}
\includegraphics[height=0.51\columnwidth]{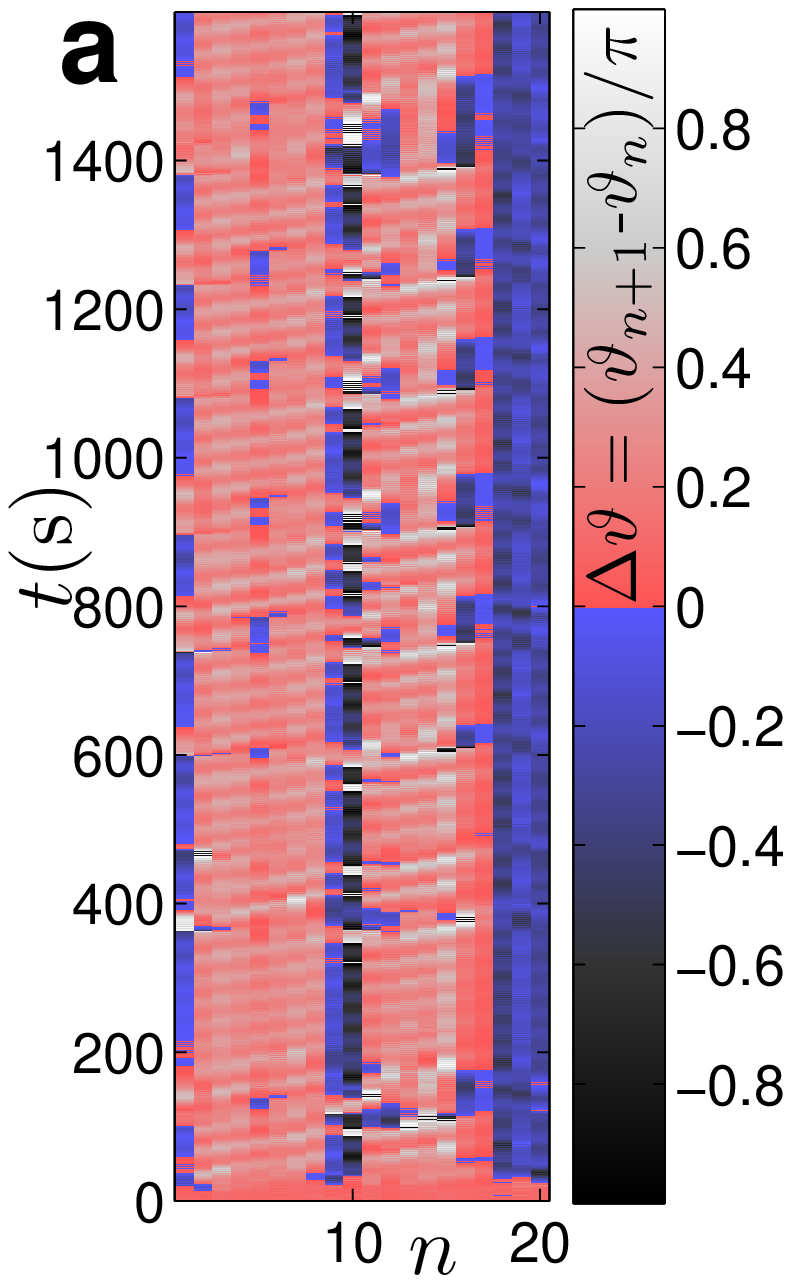}	\label{fig4a}
\includegraphics[height=0.51\columnwidth]{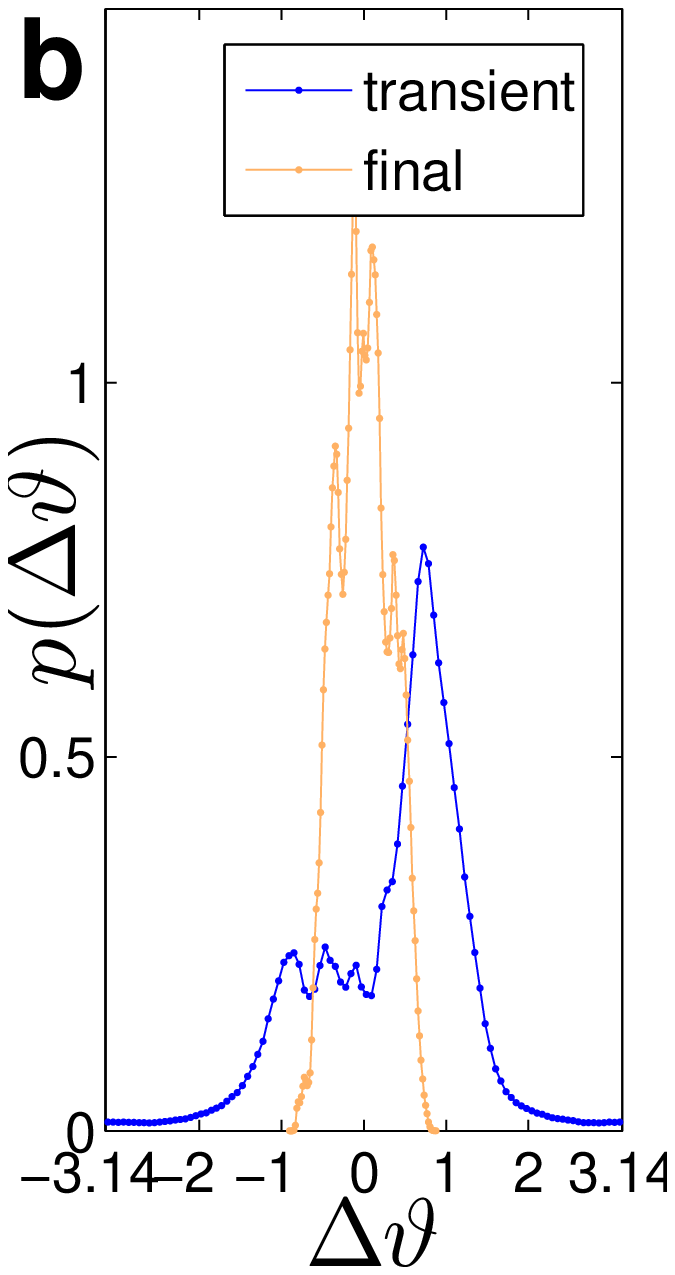}	\label{fig4b}
\includegraphics[height=0.51\columnwidth]{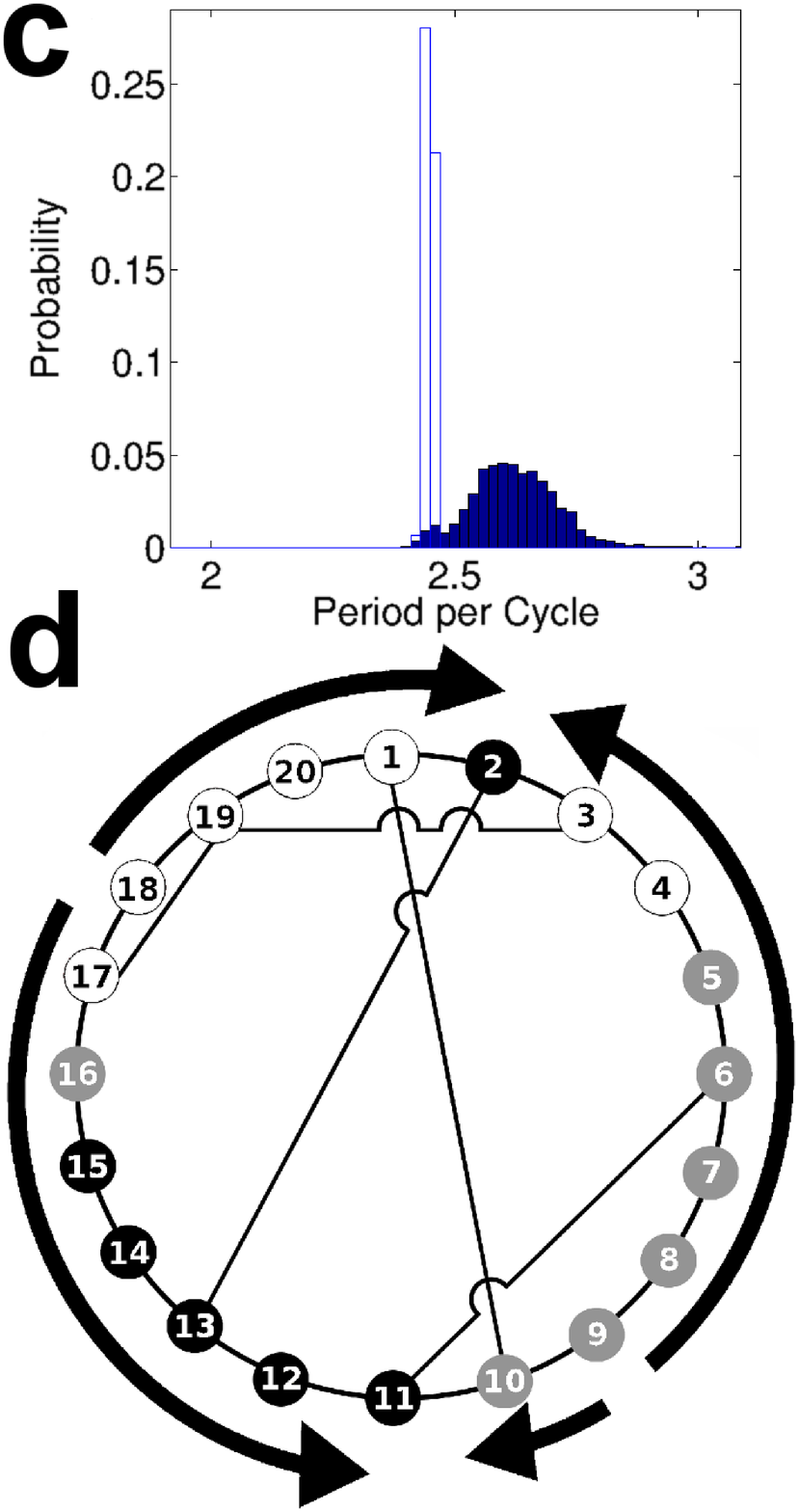}	\label{fig4cd}
\includegraphics[width=0.93\columnwidth]{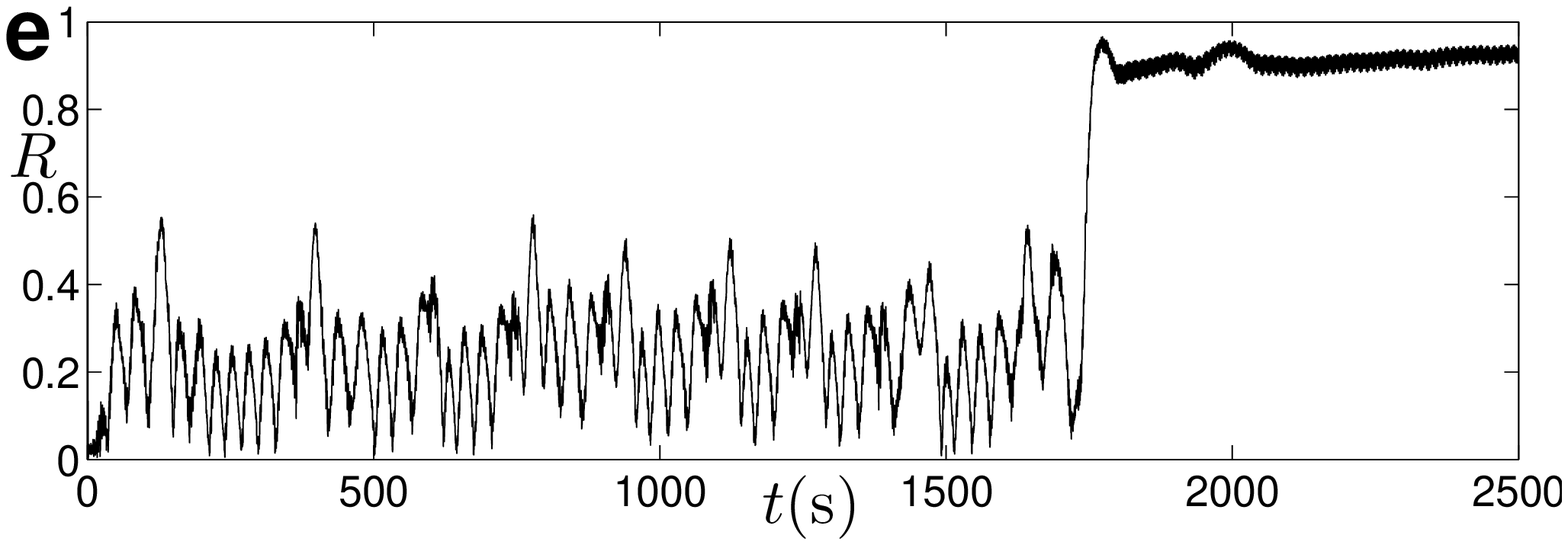}	\label{fig4e}
\end{center}
\caption{Long transient to synchronization at $\alpha=-1.3$ ($V$=1110 mV, $C\textsubscript{c}$=82 $\mu$F, $K$=0.033 mS). 
{(a)} Time evolution of next neighbor phase differences. 
{(b)} Histogram of next neighbor phase differences during the transient and the final synchronized state. 
{(c)} Histogram of the period per cycle of all electrodes (white bars are the individual periods and dark bars the transient).
{(d)} Network topology and a typical complex rotating wave pattern during the transient.  
Arrows indicate the wave direction from sources to sinks. 
{(e)} Time evolution of the Kuramoto order parameter where the arrows indicate transitions between wave patterns through phase slips.}
\label{fig-phase-turb}
\end{figure}
%
\noindent 
We have performed 16 independent trials adding cross-connections successively to a ring configuration. The
mean order parameter as a function of the added number of shortcuts is shown 
in Fig.~\ref{fig-sync-trans}c. Only three shortcuts were required for the average order parameter to exceed 0.90. 
Therefore, we can conclude that with $\alpha=0 $ a relatively small number of shortcuts 
($\sigma \approx 0.15$) induces full synchrony. When $\alpha$ was changed to -0.97, with a parallel RC coupling, we still observed rotating waves jumping over the 
connection when the first cross-connection was placed up to a distance of 5 units 
between the elements. However, when the distance was larger, instead of full synchronization, 
we observed frozen complex rotating  pattern via the formation of a source and sink pair (Fig.~\ref{fig-sync-trans}e). 
The order parameter vs. number of shortcuts graph in Fig.~\ref{fig-sync-trans}c shows that with $\alpha=-0.97$
the mean order parameter starts to increase for $n>3$, and it requires relatively 
large number ($n>6$) of random shortcuts to achieve Kuramoto order larger than 0.90.
The presence of jumping waves and complex rotating wave patterns thus contributes to resisting  complete synchronization. 
When $\alpha$ was further changed to -1.3, the trend of resisting the fully synchronized 
state continued. 
Figures \ref{fig-sync-trans}e,f demonstrate the inversion of the stationary phase profile, and thus the direction of the rotating waves and the reversal of 
source and sink, upon switching the sign of the non-isochronicity by adding a capacitance to the individual current instead of the coupling current 
\cite{Wickramasinghe:2013:062911}\footnoterecall{Note1}. 
All the patterns in Fig.~\ref{fig-sync-trans} were reproduced by phase model simulations in \footnoterecall{Note1}.
Long transients to both frozen complex wave patterns and identical synchronization were observed in experiment. 
A long transient over $1700$s in a network with five random shortcuts near $\alpha$=-1.3 is shown 
in Fig.~\ref{fig-phase-turb}. 
The time evolution of next neighbor 
phase differences during the transient is shown in Fig.~\ref{fig-phase-turb}a. At least two competing wave patterns, which are meta stable over the 
course of tens of oscillations and transform via intermittent phase slips could be observed for over 700 oscillations before the  system settled into the 
one-cluster state. 
At $\alpha=0$ the same network relaxes exponentially to synchronization in $230$s \footnoterecall{Note1}. Numerical simulations with the phase model confirmed that by changing $\alpha$ from 
0 to -1.3 the lifetime of the transient increases about 10 times and diverges as $\alpha$ approaches $|\pi/2|$ \footnoterecall{Note1}.
A density plot of next neighbor phase differences during and after the transient is depicted in 
Fig.~\ref{fig-phase-turb}b and shows an asymmetric distribution with a preferred wave length, skewed towards 
the initial left-handed rotational state. 
In addition, a wide distribution of peak to peak periods can be observed during the transient as seen in 
Fig.~\ref{fig-phase-turb}c, which marks the presence of irregular phase dynamics. A typical snapshot of a transient complex rotating wave is shown
in Fig.~\ref{fig-phase-turb}d. The arrows indicate the direction of the rotations as well 
as the sources (oscillators 9 and 18) and sinks (oscillators 2 and 11) of the 
unstable rotational wave pattern. 
The order parameter (Fig.~\ref{fig-phase-turb}e) clearly exhibits the transient behavior 
as its value changes between approximately 0.2 and 0.6 irregularly throughout 
the transient until synchronization is achieved. 
\\
In conclusion, we have observed frozen complex rotating  patterns and long transients to synchronization in electrochemical oscillations and 
numerical simulations of phase oscillators on a ring network topology with  sparse random shortcuts. Depending on the sign of non-isochronicity $\alpha$, 
either the sinks or the sources are pinned to endpoints of cross-connections in the network. At increased non-isochronicity the variability in the phase 
differences in a phase locked state decreases until synchronization is no longer possible and persistent or very long transient phase dynamics occurs. The 
presence of long transients of irregular phase dynamics could have relevance in the functioning of biological systems, e.g., in neuron dynamics where pathological 
synchronization can occur without apparent external perturbation or change of network topology. The experimentally recorded, irregular transient 
dynamics contributes to the few experimental examples of high dimensional transient chaos \cite{TelLai2008}, where system size effects on the lifetime 
of the transient irregular state could be studied.
\\
%
%
This material is based upon work supported by the National Science Foundation under Grant Number 
CHE-1465013. 
The manuscript has supplemental material \footnoterecall{Note1}.
\vspace{0.4cm}
%
%
%
\section*{\bf References} 
\bibliography{references-Saved}
%
\begin{center}
\textbf{\large Supplemental Material}
\end{center}
\setcounter{equation}{0}
\setcounter{figure}{0}
\setcounter{table}{0}
\setcounter{page}{1}
\makeatletter
\renewcommand{\theequation}{S\arabic{equation}}
\renewcommand{\thefigure}{S\arabic{figure}}

\section{Experimental Setup}
\noindent
\begin{figure}[th]\label{fig.sm01}
\begin{center}
\includegraphics[width=0.8\columnwidth]{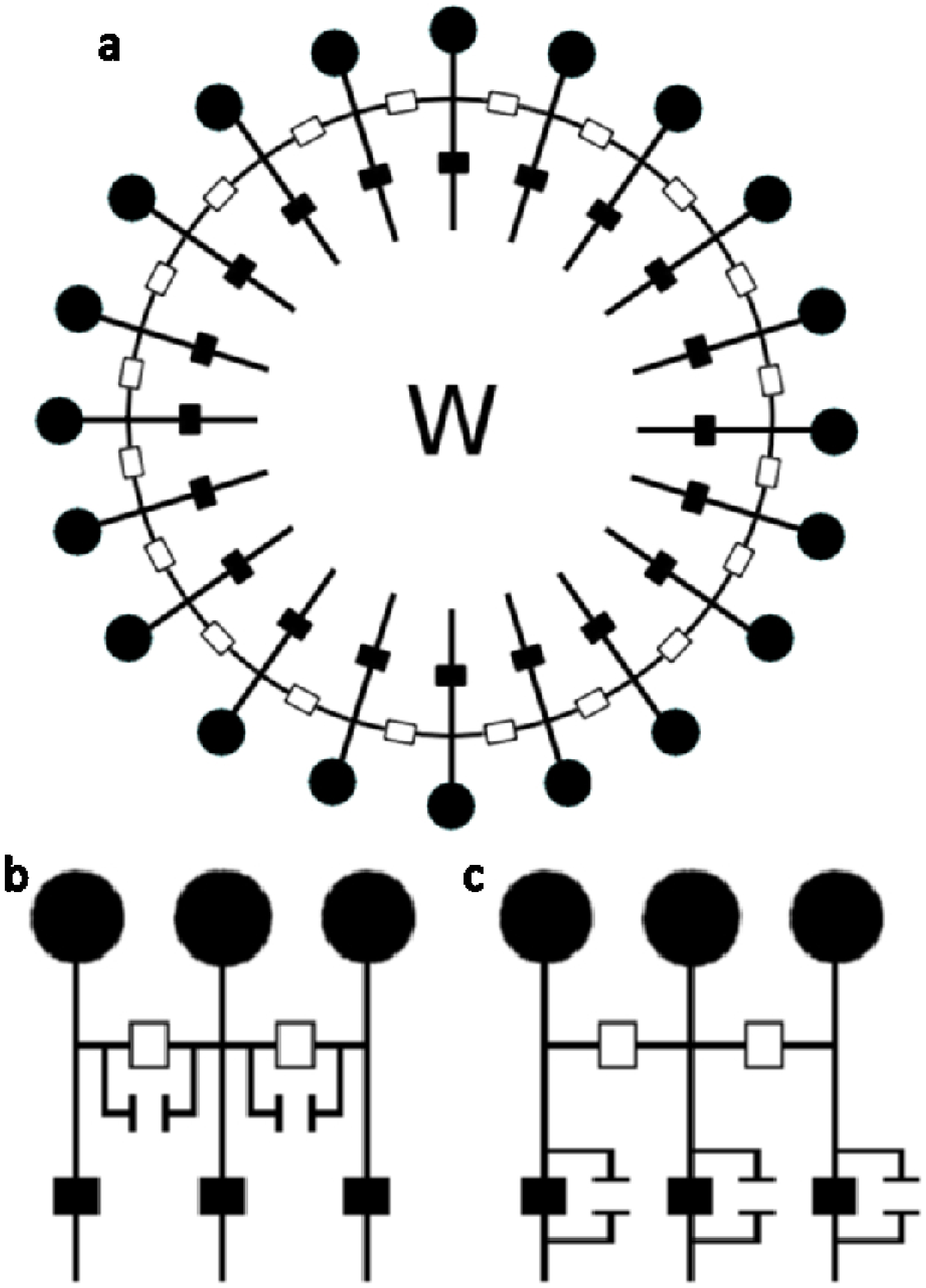}
\end{center}
\caption{Schematic of the experiments where the electrodes (black circles) are connected
 to the working point (W) of the potentiostat. $R_{c}$ are illustrated as empty squares 
 and $R_{ind}$ are illustrated by the filled squares.
{(a)} Diagram of the twenty electrode ring network with the location of $R_{c}$ and
$R_{ind}$ displayed. 
{(b)} Simplified diagram showing three electrodes with the addition of $C_{c}$.
{(c)} Simplified diagram showing three electrodes with the addition of $C_{ind}$.}
\label{si-exp-schem}
\end{figure}
Experiments are performed by a standard electrochemical cell with
3 M H\textsubscript{2}SO\textsubscript{4} as the electrolyte. The
counter electrode is a platinum coated titanium rod, the reference
electrode is Hg/Hg\textsubscript{2}SO\textsubscript{4}/sat. K\textsubscript{2}SO\textsubscript{4},
and the working electrode is an array of 20 nickel, 1.00 mm diameter,
wire cross sections embedded in epoxy. The temperature is held at 10 \textsuperscript{o}C. Each nickel
electrode in the array is connected to the potentiostat (ACM Instruments
GillAC) through an individual resistance $R_{ind}$. 
The frequencies of the electrodes are controlled by an increase or decrease of 
$R_{ind}$ and regulated to $0.4$Hz $\pm 2.5$mHz at  $R_{ind}=1k\Omega$ in the beginning of the experiment. 
The system is coupled locally into a ring topology by coupling resistance
($R_{c}$) as shown in  Fig.\ ~\ref{si-exp-schem}a. A constant
potential ($V$) is applied and the oscillatory current of each electrode
in the array is measured at 200Hz by a potential drop over $R_{ind}$. 
The coupling strength $K$ is reported as the inverse $1/R_{c}$ of the coupling resistance. Non-isochronicity is introduced into the network by the addition of 
capacitance C. The non-isochronticity is negative for a capacitance $C_c$ parallel to $R_{c}$ and positive for a capacitance $C_{ind}$ parallel to $R_{ind}$ as seen in 
Fig.~\ref{si-exp-schem}b,c. The level of nonisochronicity induced by the 
capacitance in a pair of ocillators was studied in great detail in a previous 
publication \cite{Wickramasinghe:2013:062911}.
\section{Method}
\noindent
\begin{figure}[th]\label{fig.sm02}
\begin{center}
\includegraphics[width=0.9\columnwidth]{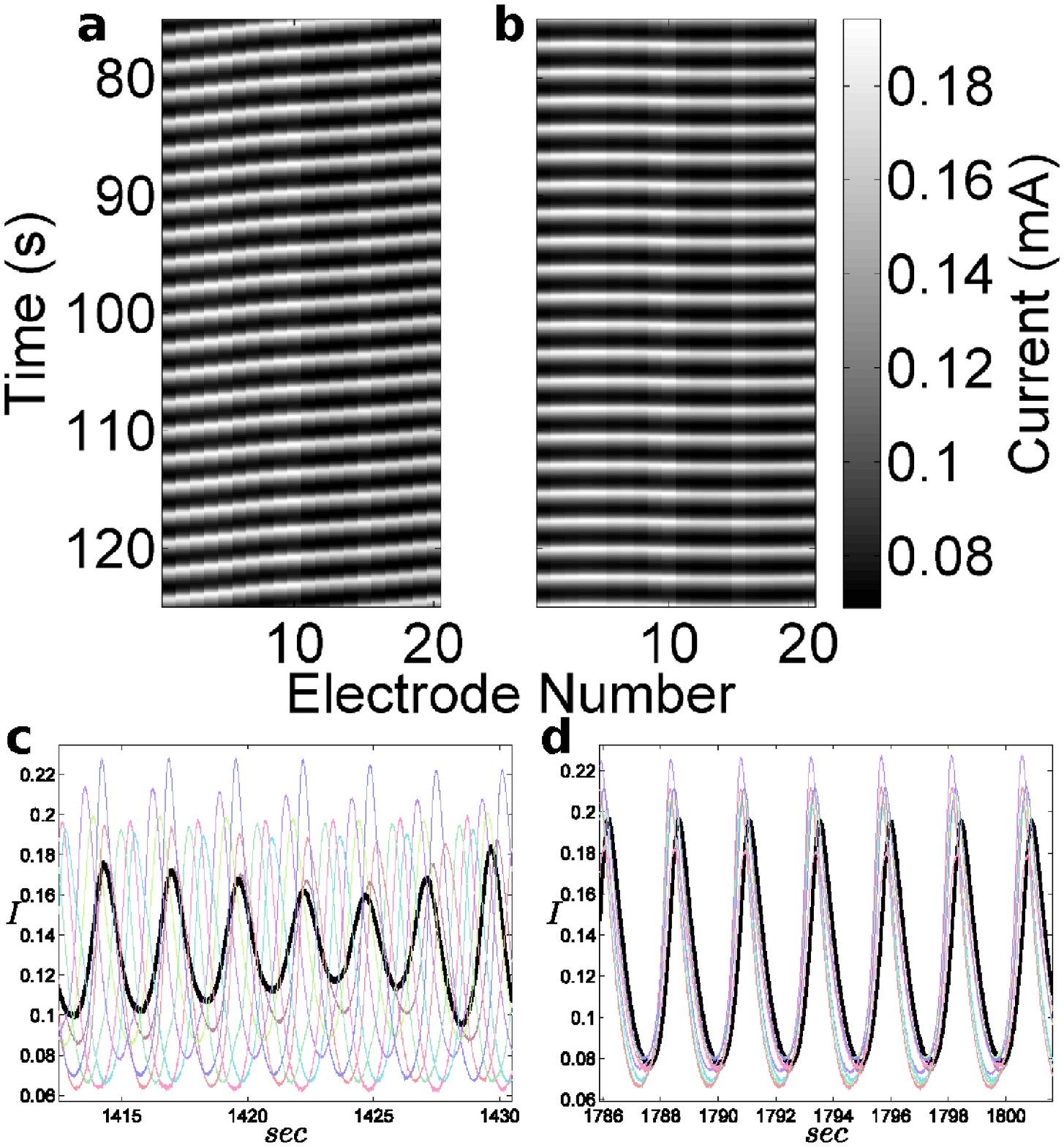}
\end{center}
\caption{Current oscillation of electrodes.
{(a)} Space-time plot of the electrode currents on the ring network in a phase-locked rotational wave state with no shortcus 
($V$= 1105 mV, $K$= 0.20 mS, $\alpha$= 0). 
{(b)} Space-time plot of the electrode currents on the  ring network with in-phase synchronization with a 1-11 cross connection.
{(c)} Measured currents of cross-cut oscillators (1,2,3,6,10,11,13,17,19) showing typical irregular transient behavior ($V$= 1110 mV, $C_c$= 82$\mu$F $K$= 0.033 mS, $\alpha$= -1.3).
{(d)} Measured currents of cross-cut oscillators (1,2,3,6,10,11,13,17,19) exhibiting in-phase synchronization.}
\label{si-osc-data}
\end{figure}
\begin{figure}[th]\label{fig.sm03}
\begin{center}
\includegraphics[width=0.9\columnwidth]{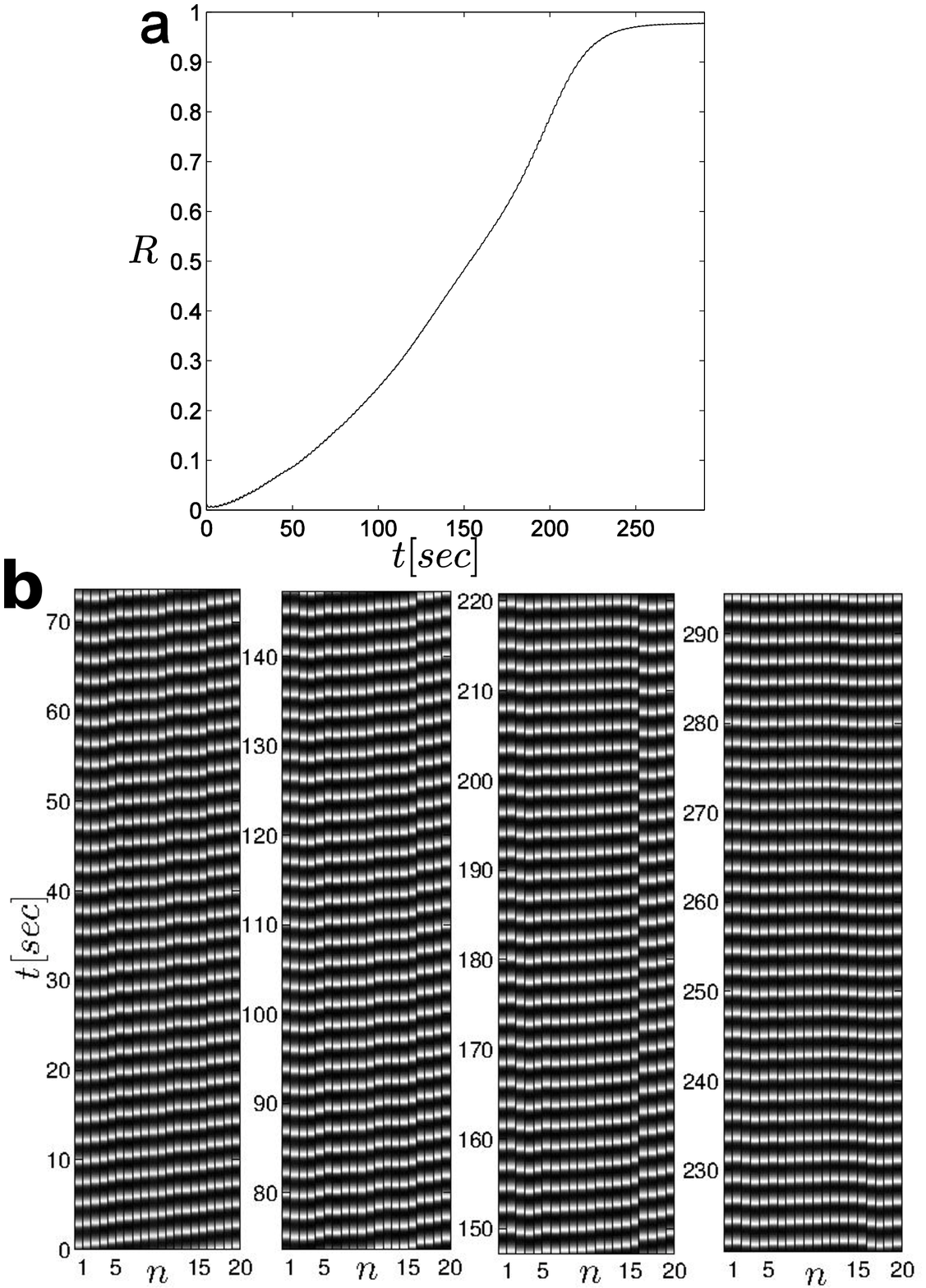}
\end{center}
\caption{Short transient to synchronization at $\alpha$ = 0 ($V$= 1105 mV, $K$= 0.033 mS).
{(a)} Time evolution of the Kuramoto order parameter.
{(b)} Space-time plot of the current during the transient from the initial rotational state to in-phase synchronization.}
\label{short-trans}
\end{figure}
Once coupled into the ring topology, global negative feedback is applied
to linearly increase the phase difference between neighboring oscillators according to 
$\vartheta_{n}=2\pi n/N$. The phase pattern is stable when feedback is removed 
(see Fig.~\ref{si-osc-data}a). Additional connections are added randomly into 
the system with the same $R_{c}$ and $C_{c}$ value as the existing connections 
in the network. 
If the system did not synchronize in-phase during the addition of the connection, a second 
connection is chosen by the same method as the first and added to the network.
The process is repeated until in-phase synchronization is observed. Further additions of shortcuts are assumed to have no desynchronizing effect. The non-rotating, synchronized 
state is shown as a space-time plot in Fig.~\ref{si-osc-data}b. Transient dynamics to a synchronized state may be exponential relaxation (Fig.\ref{short-trans}) or irregular phase dynamics with periods and amplitudes fluctating in time (Fig.~\ref{si-osc-data}c). In synchrony, the period is identical for all 
oscillators and the amplitudes are much more homogeneous as shown in Fig.~\ref{si-osc-data}d.
\section{Determination of Alpha}
\noindent
\begin{figure}[th]\label{fig.sm04}
\begin{center}
\includegraphics[width=0.49\columnwidth]{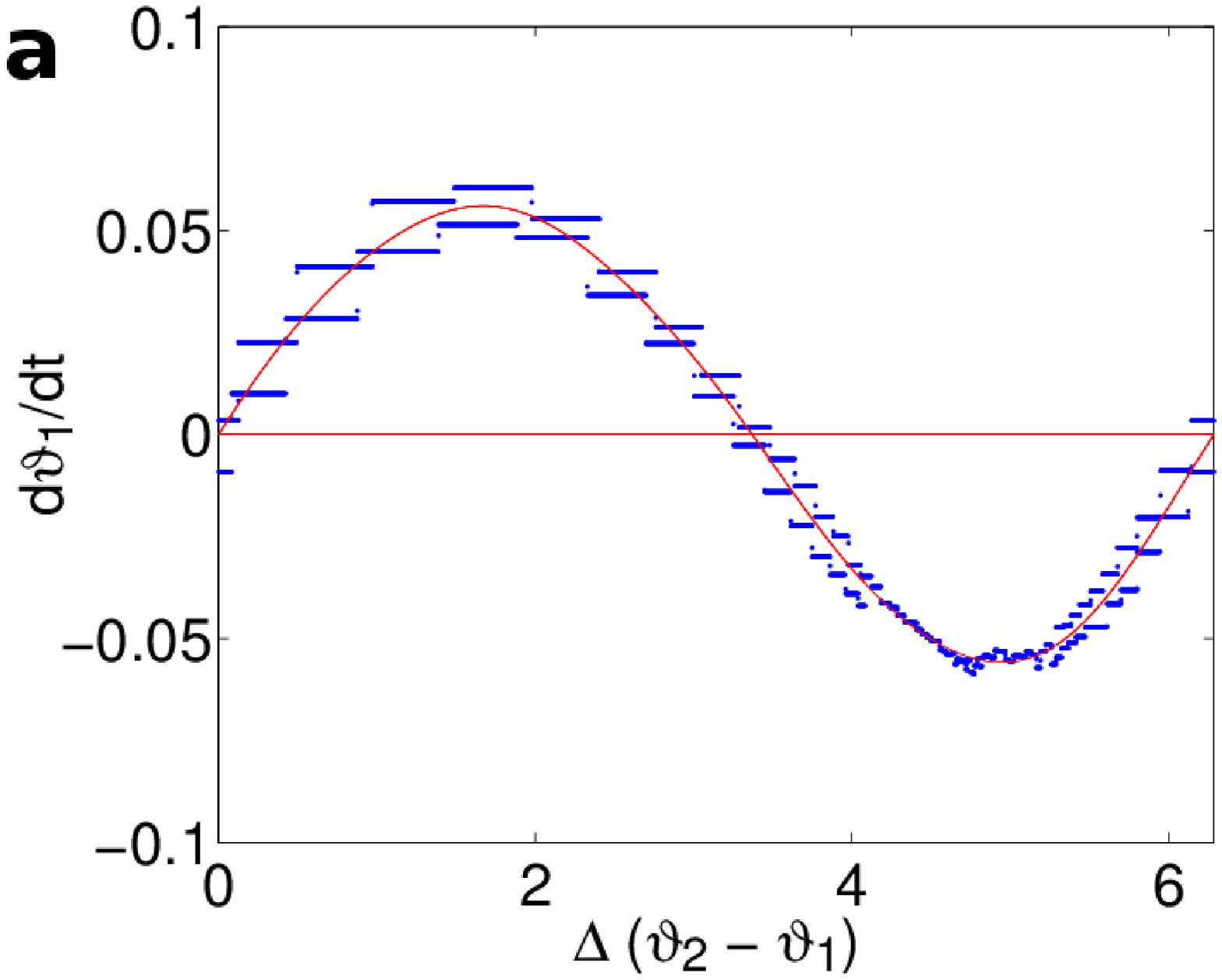}
\includegraphics[width=0.49\columnwidth]{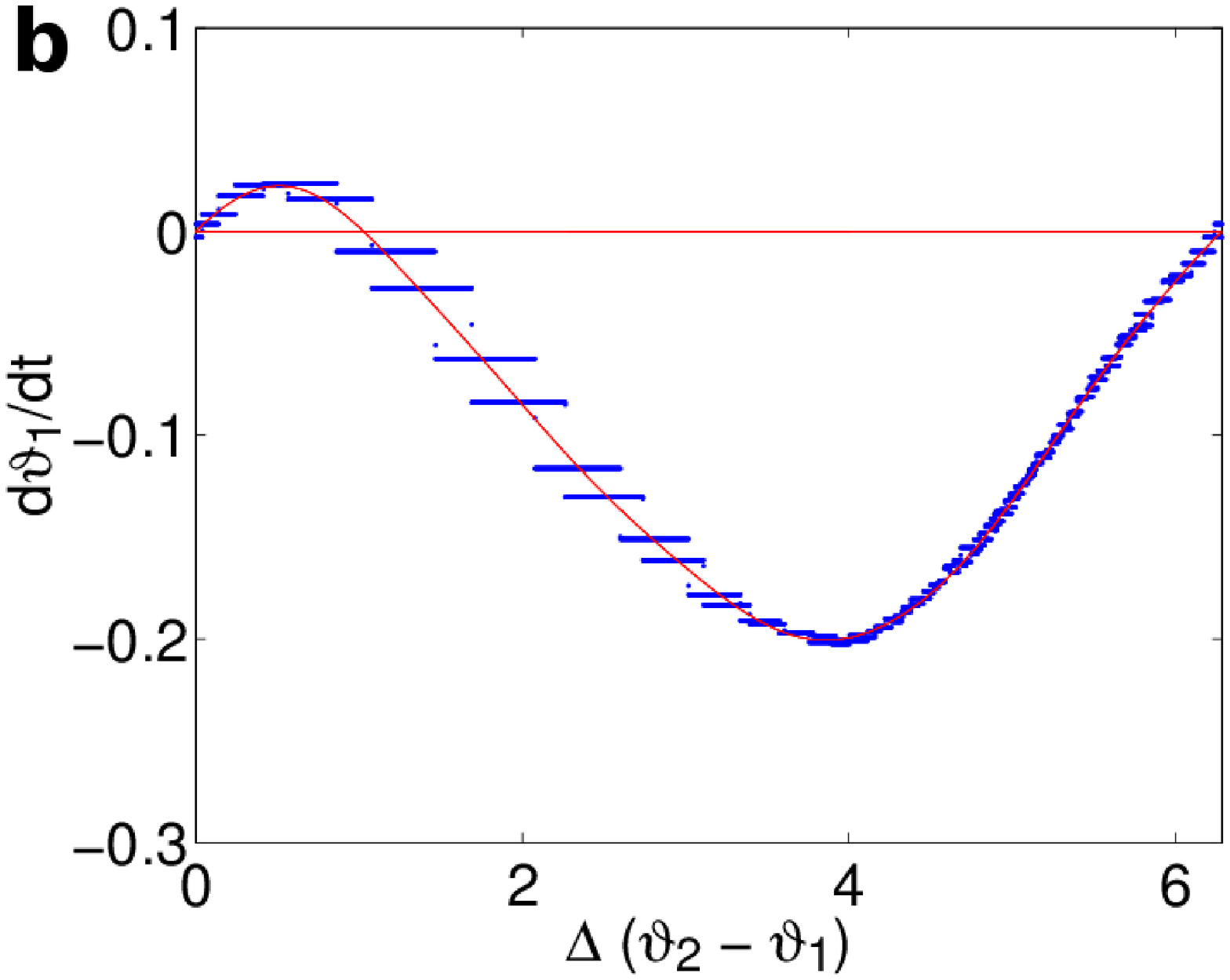}
\includegraphics[width=0.49\columnwidth]{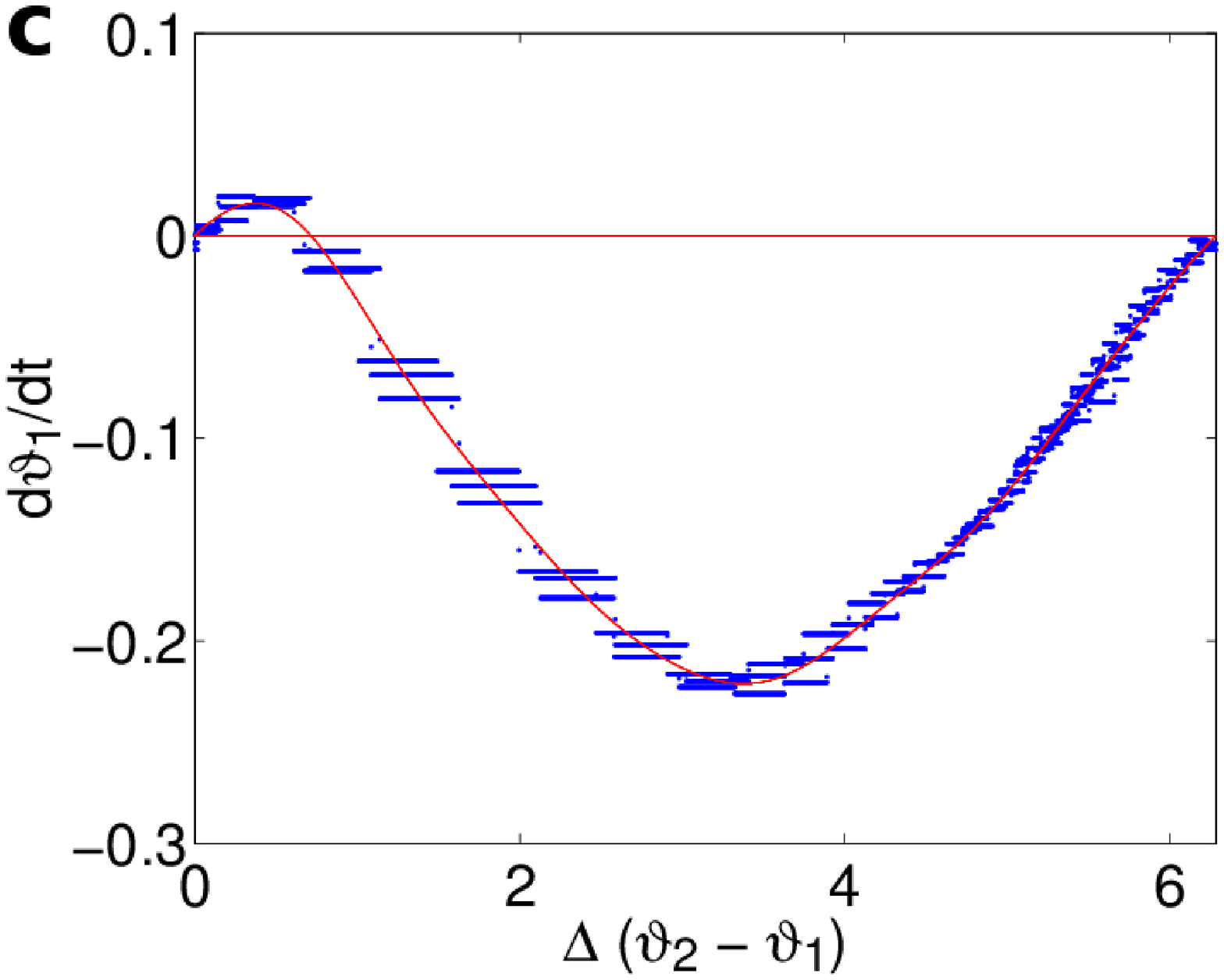}
\end{center}
\caption{Experimentally measured interaction functions ($V$=1105 mV, $R_{ind1}$=1 $k\Omega$, $R_{ind2}$=900 $\Omega$).
{(a)} $\alpha$=0 ($R_{c}$= 20 $k\Omega$). 
{(b)} $\alpha$=-0.97 ($R_{c}$= 23 $k\Omega$, $C_{c}$= 35.65 $\mu F$).
{(c)} $\alpha$=-1.3 ($R_{c}$= 69 $k\Omega$, $C_{c}$= 47.54 $\mu F$).}
\label{si-int-fun}
\end{figure}
The experimental value of non-isochronicity is determined from the interaction function of 
two coupled oscillators with a frequency difference of approximately 20 mHz, achieved by 
an $R_{ind}$ difference of 100 $\Omega$. In order to examine the interaction function, 
the coupling strength is lowered from the value of the experiments such that phase 
slips are present. The product of $R_{c}$ and $C_{c}$ is kept constant to maintain 
the same non-isochronicity as in the experiments. The interaction function is obtained by 
plotting the derivative of the phase of an oscillator (minus its natural frequency) as a 
function of the phase difference between the coupled oscillators. More details on the 
procedure is given in a previous publication \cite{Wickramasinghe:2013:062911}. The 
experimentally obtained interaction functions are shown in Fig.~\ref{si-int-fun}a,b,c. 
The value of the sine and cosine harmonic coefficients of the interaction function are 
used to calculate the non-isochronicity (or phase shift parameter):
\begin{equation}
\alpha = \arctan\left(\frac{A_{1}}{B_{1}}\right)\,
\label{eq:alpha}
\end{equation}
\noindent where $A_{1}$ is the  sine coefficient and $B_{1}$ is the cosine coefficient of 
the first harmonic of the interaction function obtained by fast fourier transform. The same procedure 
is used for positive $\alpha$ with the product of $R_{ind}$ and $C_{ind}$ held constant.
\section{20 Phase Oscillators}
\noindent
\begin{figure}[th]\label{fig.sm05}
\begin{center}
\includegraphics[width=0.85\columnwidth]{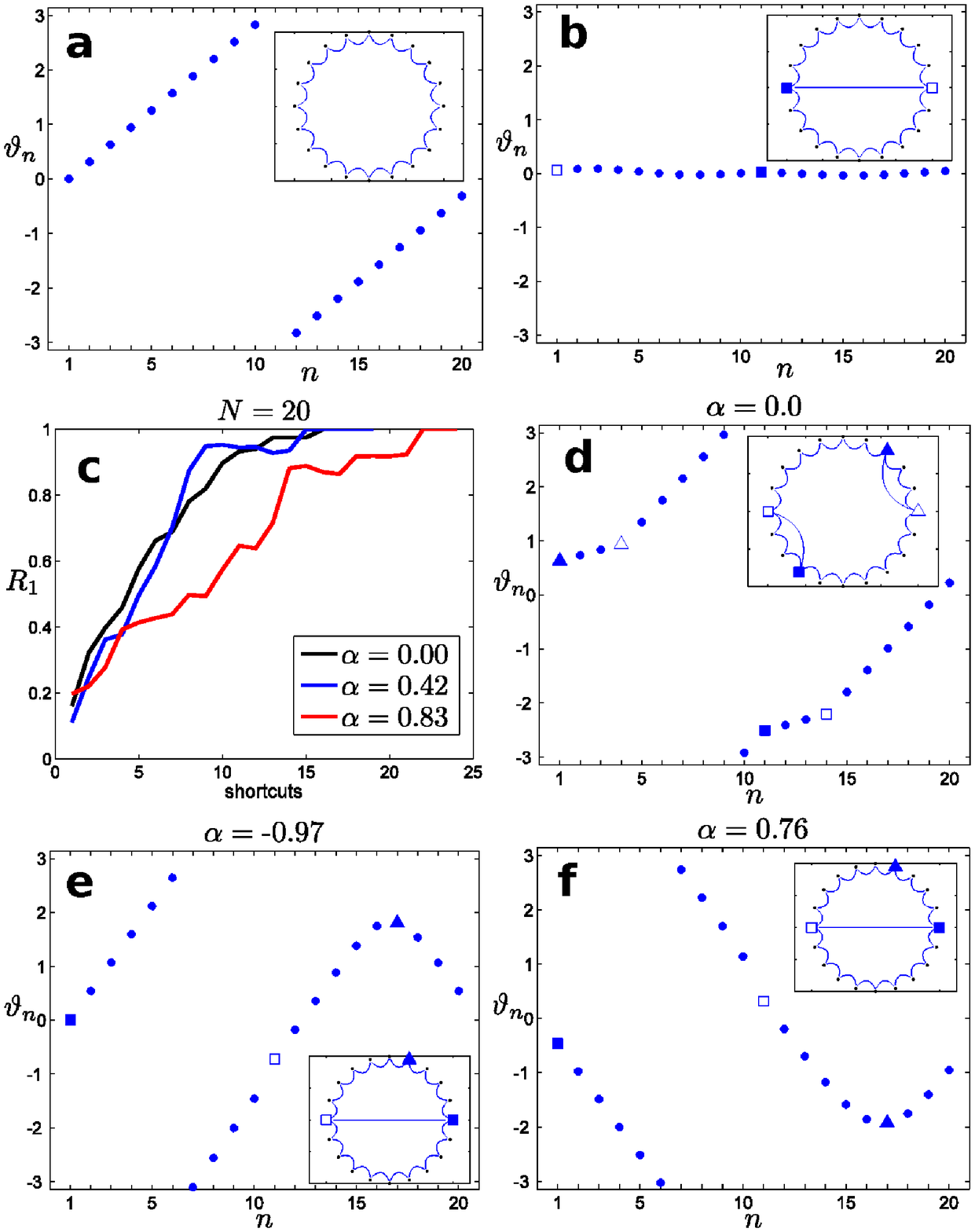}
\end{center}
\caption{Dynamics of $N$= 20 phase oscillators.
{(a)} Rotating wave phase simulation, ($\alpha$= 0).
{(b)} Synchronized state in the presence of a 1-11 cross connection.
{(c)} Mean order parameter after transient as a function of sequentially added shortcuts.
{(d)} Jumping wave phase simulation, ($\alpha$= 0).
{(e)} Phase simulation of frozen complex rotating wave pattern with a source (triangle) away 
from the shortcut heterogeneity, ($\alpha$= -0.97).
{(f)} Phase simulation of frozen complex rotating wave pattern with a sink (triangle) away 
from the shortcut heterogeneity, ($\alpha$= 0.76).}
\label{si-fig3sim}
\end{figure}

\begin{figure}\label{fig.sm06}
\begin{center}
\includegraphics[width=0.8\columnwidth]{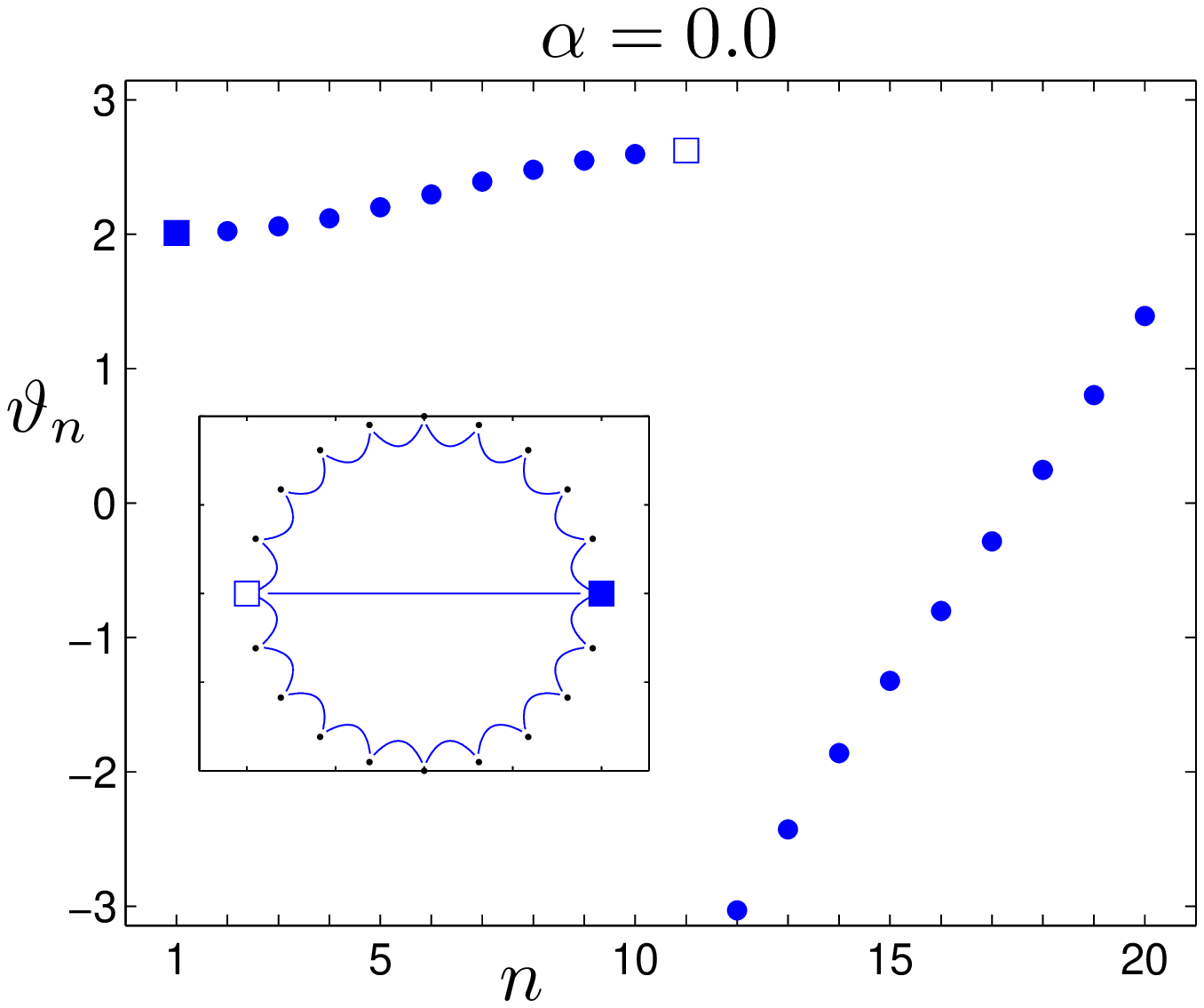}
\end{center}
\caption{At $\alpha=0$ with a single connection between oscillator 1 and 11 in a ring of $N=20$ phase oscillators, and a slightly perturbed rotating wave initial condition a jumping wave accross one part of the ring and a rotating wave accross the other part of the ring forms. This configuration is not observed in the experiments which synchronize completely.}
\label{si-deviate}
\end{figure}

\begin{figure}\label{fig.sm07}
\begin{center}
\includegraphics[width=0.493\columnwidth]{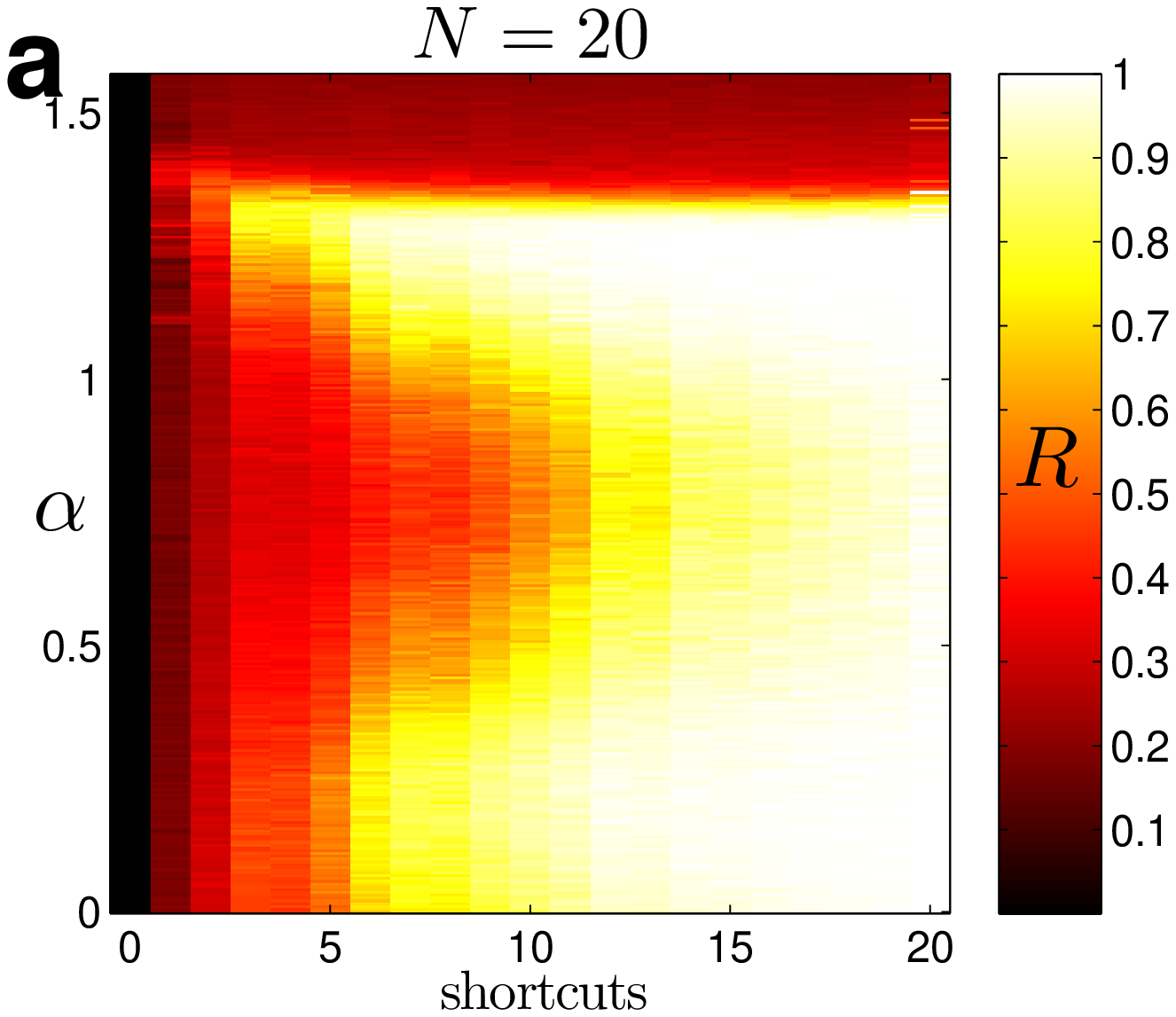}
\includegraphics[width=0.493\columnwidth]{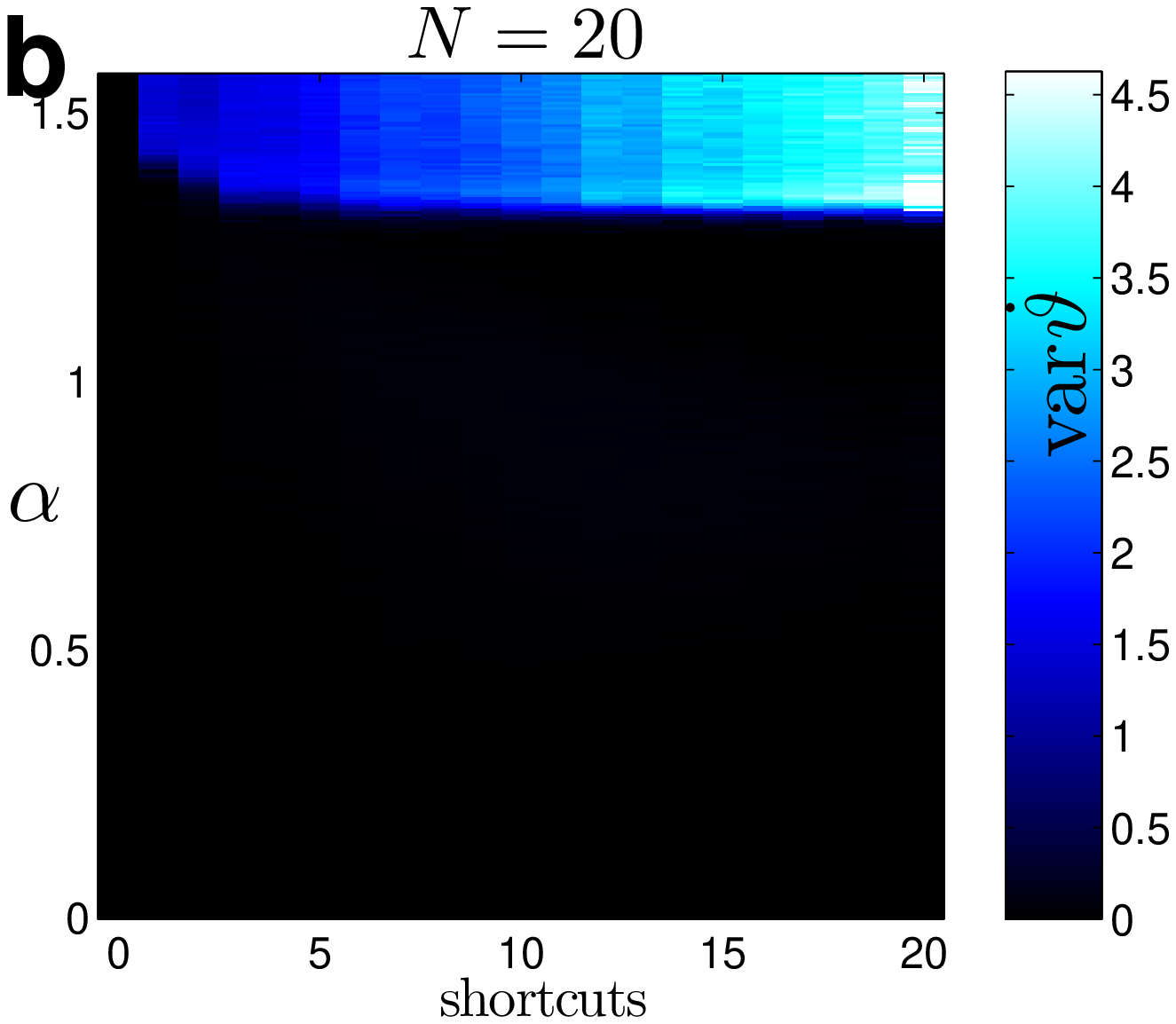}
\end{center}
\caption{(a) Order parameter and (b) variance of phase velocities averaged over 200 realizations of random networks with $N$=20 phase oscillators as a 
function of $\alpha$ and the number of shortcuts. Initial conditions are a rotating wave of winding number one. The data was evaluated after reaching a phase locked state or at most $t=500$ time units.}
\label{si-alphsc}
\end{figure}

\begin{figure}\label{fig.sm08}
\begin{center}
\includegraphics[width=0.7\columnwidth]{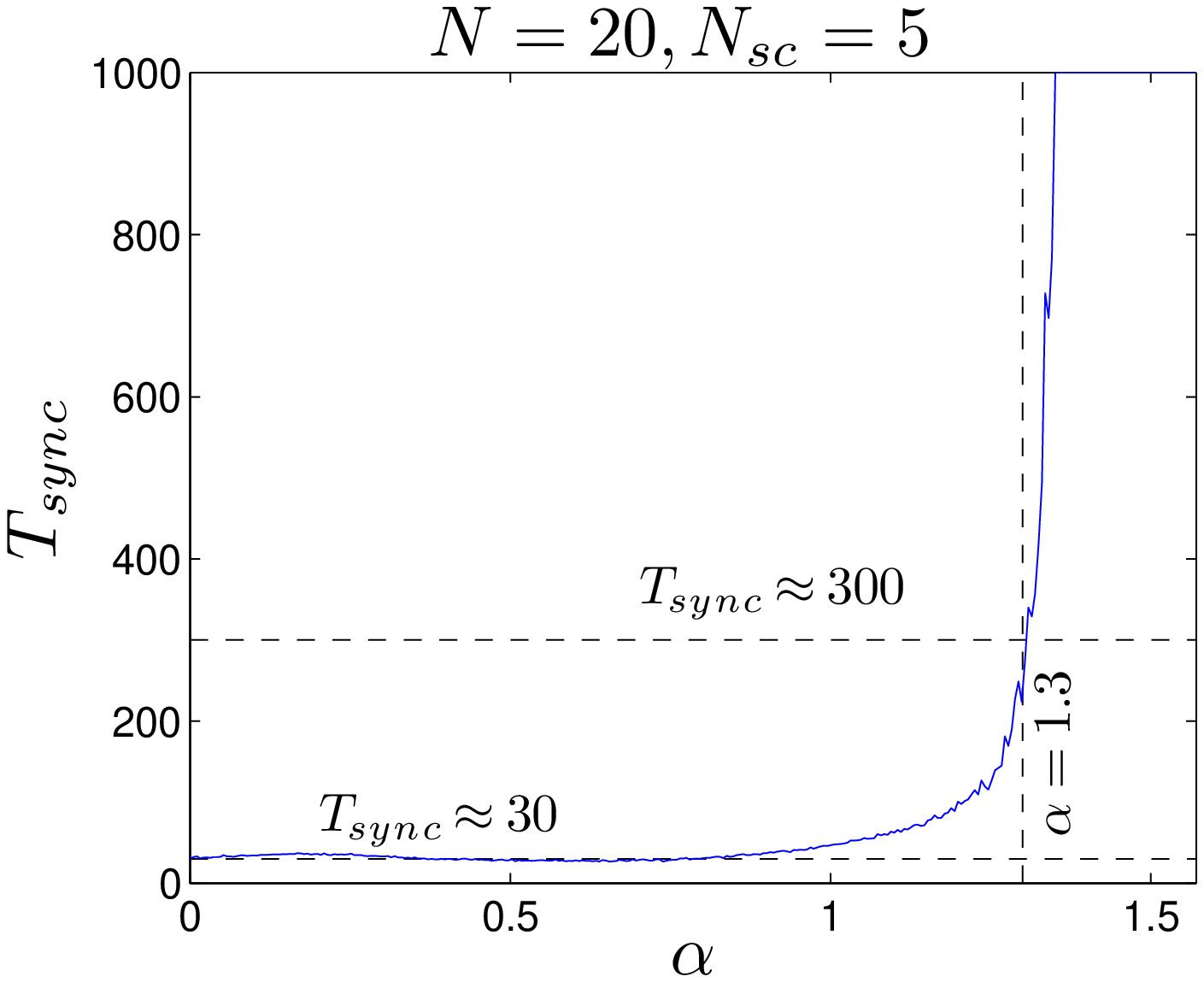}
\end{center}
\caption{Median time to phase locking as a function of $\alpha$ 
averaged over 200 realizations, for an $N$= 20 system with five random shortcuts.}
\label{si-transtime}
\end{figure}

While the main text \cite{Sebek2016} provides simulations for systems with $N$= 500 phase oscillators, 
simulations with $N$= 20 phase oscillators were performed to demonstrate the qualitative agreement between
experiments and the phase model. Figure 3 of the main text \cite{Sebek2016} is reproduced with 
phase oscillators using the same method as in the experiments. A stable 
rotational state (Fig.~\ref{si-fig3sim}a) is found when there are no shortcuts and 
$\alpha$=0. The stable state of complete synchronization exists with a connection between oscillator 1 and 11 (Fig.~\ref{si-fig3sim}b).
However, in contrast to the experiments this state is not reached from an initial rotating wave. Instead a slight break of the symmetry in the initial rotating wave will 
lead to in-phase synchronization on one side of the ring (jumping wave) and a rotating wave on the other part of the ring (Fig.~\ref{si-deviate}).
If $\alpha$ is increased, the number of shortcuts required to achieve synchrony increases as seen 
in the plot of the mean order parameter as a function of the number of shortcuts shown in Fig.~\ref{si-fig3sim}c and Fig.~\ref{si-alphsc}a. A larger number of short cuts than in the experiments is needed to achieve complete synchronization in the phase model. We attribute this to the lack of the amplitude degree of freedom in the phase model which can increase the basin of attraction of frozen complex rotating wave patterns.
Jumping waves (Fig ~\ref{si-fig3sim}d) form with the addition of
cross-connections over small distances at $\alpha$= 0. If $\alpha$ is sufficiently negative a wave source 
and a sink form with the source position away from the cross-connection network heterogeneity 
whereas for sufficiently positive $\alpha$ a wave source and sink form in opposing 
positions of negative $\alpha$ (Fig.~\ref{si-fig3sim}e,f).
The typical system states of frozen complex wave patterns, complete synchronization and irregular phase dynamics
shown in Figure 1 in the main text \cite{Sebek2016} are also observed with as few as $N$=20 phase oscillators.  Fig.~\ref{si-alphsc}a,b show the ensemble averaged Kuramoto order parameter and the variance of phase velocities as a function of the number of short cuts and non-isochronicity $\alpha$ for 200 random network realizations.
The transient time median to a synchronized state where the difference between the maximal and the minimal phase velocity is less than $10^{-3}$ in ring networks of $N$=20 phase oscillators with 
$N_{sc}$=5 short cuts as a function of $\alpha$ is shown in Fig.~\ref{si-transtime}. Our simulations show a divergence of this transient time as $\alpha$ is increased. At low $\alpha$ in the regime of frozen complex wave patterns the transient time median is approximately constant but increases by a factor of 10 at $\alpha=1.3$. This behavior is consistent with the experimentally observed long and irregular transient at $\alpha=-1.3$ which with $T_{trans} \approx 1700s$ (\cite{Sebek2016} Fig.4e) is about 7 times longer than the exponential relaxation with $T_{trans} \approx 230s$ for the same network and the same coupling strength $K$=0.033 mS at $\alpha=0$ (Fig.~\ref{short-trans}a). 

\section{Clustering}
\begin{figure}[th]\label{fig.sm09}
\begin{center}
\includegraphics[width=0.8\columnwidth]{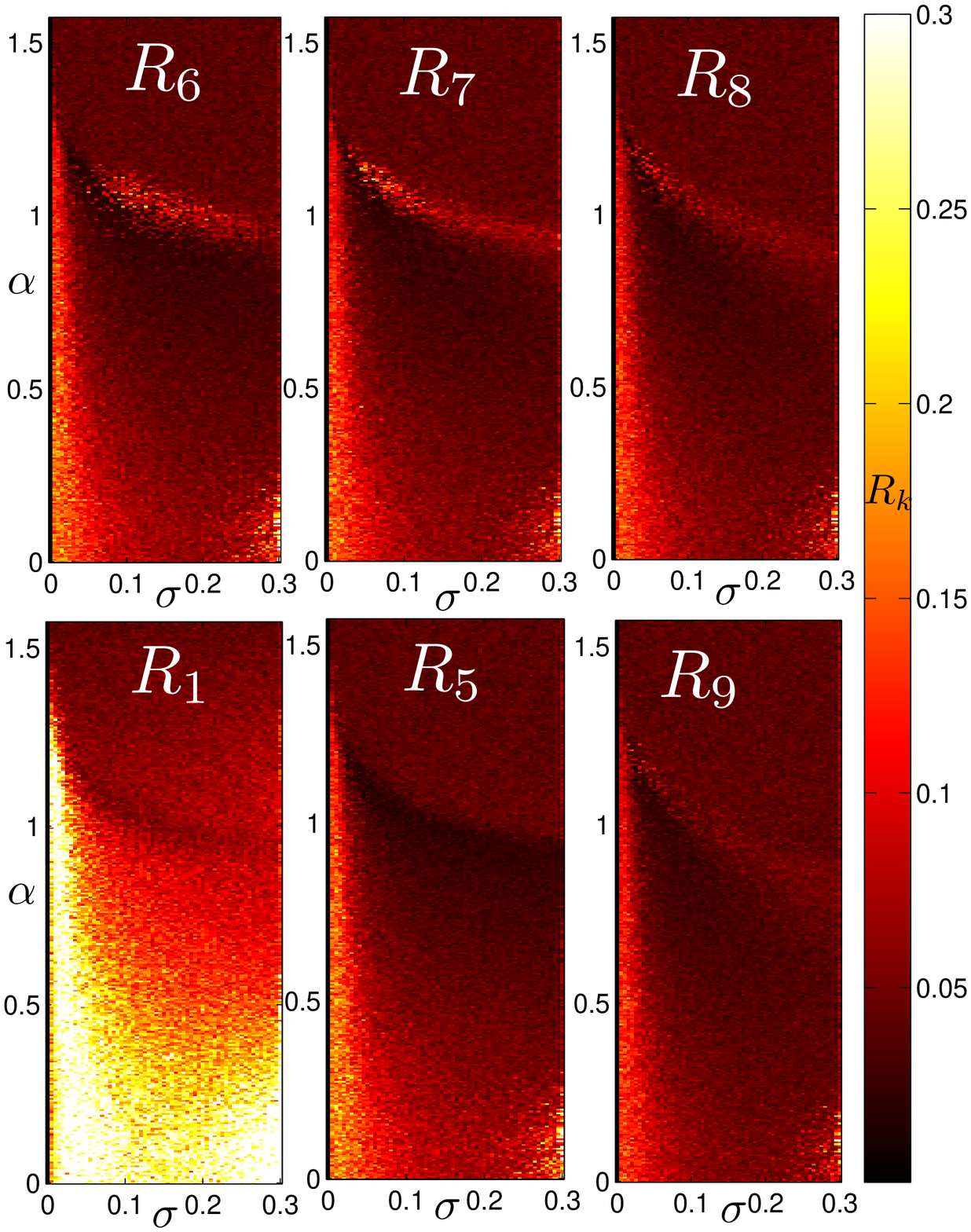}
\end{center}
\caption{k-cluster order parameters $R_k$ - $R_6$, $R_7$, $R_8$, $R_1$, $R_5$ and $R_9$ 
as functions of shortcut density, $\sigma$, and non-isochronicity $\alpha$, for $N$= 500 
oscillators, averaged over 10 network realizations at $t$= 500 time units.}
\label{si-clusters}
\end{figure}

\noindent
Clustering is observed for networks of identical phase oscillators with shortcut densities between $\sigma=0.05$ and $0.2$ in a narrow parameter region along the critical line between
the regions of frozen complex wave patterns and irregular phase dynamics. We have not observed clustering in the experiments. The reasons for that might be the small number of oscillators, small heterogeneities in the frequencies, amplitude effects or simply not enough observations. Clustering in the simulations with $N=500$ phase oscillators is evident in the cluster order parameters shown in in Fig.~\ref{si-clusters}. Figure 1d in the main text \cite{Sebek2016} shows $R_6$ and $R_7$ of the same data with restricted shortcut density $\sigma=0.05$. In our simulations of the Kuramoto phase equations on our random network model we have observed clustering in networks as large as $N=10^5$ oscillators and $N_{sc}=5000$ shortcuts. In the clustering region the cluster number and strength of the clustering a system may display is distributed with a dependence on the position on the critical line. Shown in Fig.~\ref{si-clusters} are the average cluster order parameters. Individual systems may demonstrate stronger or weaker clustering, a different cluster number or no clustering at all.
%
\end{document}